\documentclass[nonacm,sigplan]{acmart}

\usepackage[]{hyperref}

\usepackage{amsmath,amsfonts}
\usepackage{algorithmic}
\usepackage{booktabs}
\usepackage{graphicx}
\usepackage{textcomp}
\usepackage{xcolor}
\usepackage{fancyhdr}
\usepackage{url}
\usepackage{microtype}
\usepackage{framed}
\usepackage{endnotes}

\def\BibTeX{{\rm B\kern-.05em{\sc i\kern-.025em b}\kern-.08em
    T\kern-.1667em\lower.7ex\hbox{E}\kern-.125emX}}

\usepackage{pifont}
\usepackage{graphicx}
\usepackage{xcolor}

\newcommand{\rvs}[1]{#1}
\newcommand{\blue}[1]{{\color{blue}#1}}
\newcommand{\red}[1]{{\color{red}#1}}

\newcommand{\diff}[1]{}

\newcommand{\nocmt}{1}
\newcommand{\stale}{1}

\newcommand{\name}{\textsc{CtXnL}}
\newcommand{\vanilla}{CXL-vanilla}

\newcommand{\cxl}{CXL}
\newcommand{\cxlio}{\texttt{CXL.io}}
\newcommand{\cxlmem}{\texttt{CXL.mem}}
\newcommand{\cxlcache}{\texttt{CXL.cache}}
\newcommand{\cxlbi}{\texttt{CXL.BI}}

\newcommand{\oone}{\ding{172}}
\newcommand{\ttwo}{\ding{173}}

\newcommand{\onee}{\ding{182}}
\newcommand{\twoo}{\ding{183}}

\begin{document}
\title{Enabling Efficient Transaction Processing on CXL-Based Memory Sharing}


\author{Zhao Wang}
\affiliation{%
  \institution{School of Integrated Circuits}
  \institution{School of Computer Science}
  \institution{Peking University}
  \city{Beijing}
  \country{China}
}
\email{wangzhao21@pku.edu.cn}

\author{Yiqi Chen}
\affiliation{%
  \institution{School of Integrated Circuits}
  \institution{Peking University}
  \city{Beijing}
  \country{China}
}
\email{yiqi.chen@pku.edu.cn}

\author{Cong Li}
\affiliation{%
  \institution{School of Integrated Circuits}
  \institution{Peking University}
  \city{Beijing}
  \country{China}
}
\email{leesou@pku.edu.cn}

\author{Yijin Guan}
\authornote{co-corresponding author}
\affiliation{%
  \institution{DAMO Academy, Alibaba Group}
  \institution{Hupan Lab}
  \city{Hangzhou}
  \country{China}
}
\email{yijin.gyj@alibaba-inc.com}

\author{Dimin Niu}
\affiliation{%
  \institution{DAMO Academy, Alibaba Group}
  \institution{Hupan Lab}
  \city{Hangzhou}
  \country{China}
}
\email{dimin.niu@alibaba-inc.com}

\author{Tianchan Guan}
\affiliation{%
  \institution{DAMO Academy, Alibaba Group}
  \institution{Hupan Lab}
  \city{Hangzhou}
  \country{China}
}
\email{tianchan.gtc@alibaba-inc.com}

\author{Zhaoyang Du}
\affiliation{%
  \institution{DAMO Academy, Alibaba Group}
  \institution{Hupan Lab}
  \city{Hangzhou}
  \country{China}
}
\email{zhaoyang.dcy@alibaba-inc.com}

\author{Xingda Wei}
\affiliation{%
  \institution{Institute of Parallel and Distributed Systems, SEIEE}
  \institution{Shanghai Jiao Tong University}
  \city{Shanghai}
  \country{China}
}
\email{wxdwfc@sjtu.edu.cn}

\author{Guangyu Sun}
\authornotemark[1]
\affiliation{%
  \institution{School of Integrated Circuits}
  \institution{Beijing Advanced Innovation Center for Integrated Circuits}
  \institution{Peking University}
  \city{Beijing}
  \country{China}
}
\email{GSun@pku.edu.cn}

\begin{sloppypar}
\begin{abstract}


Transaction processing systems are the crux for modern data-center applications, yet current multi-node systems are slow due to network overheads. This paper advocates for Compute Express Link (CXL) as a network alternative, which enables low-latency and cache-coherent shared memory accesses. However, directly adopting standard CXL primitives leads to performance degradation due to the high cost of maintaining cross-node cache coherence. To address the CXL challenges, this paper introduces \name, a software-hardware co-designed system that implements a novel hybrid coherence primitive tailored to the loosely coherent nature of transactional data. The core innovation of \name~is empowering transaction system developers with the ability to selectively achieve data coherence. Our evaluations on OLTP workloads demonstrate that \name~enhances performance, outperforming current network-based systems and achieves with up to 2.08x greater throughput than vanilla CXL memory sharing architectures across universal transaction processing policies.



\end{abstract}


\maketitle 
\pagestyle{plain} 

\section{Introduction}  \label{sec:intro}

Transactions provide high availability and strict serializability, simplifying programming and reasoning about concurrency issues. However, scaling single-node transaction systems~\cite{silo_sosp13, abyss_vldb14, polyjuice_osdi21} beyond a node is challenging due to their reliance on shared-memory architectures. When datasets exceed a single node’s memory capacity, a common solution is to employ a distributed transaction system that partitions the data into multiple shards and distributes them across different nodes. This approach utilizes networking techniques, such as RDMA~\cite{farm_nsdi14, compromise, drtmh, fasst}, to fetch remote data and synchronize transactions across nodes. However, distributed transactions are notorious for their poor performance, often attributed to network stack overheads~\cite{farm_nsdi14, compromise, drtmh, fasst, grappa_atc15, guideline_atc16}. Despite numerous optimizations in software~\cite{fasst, drtmh, grappa_atc15, guideline_atc16} and hardware~\cite{ddio_ispass20, ddio-doc, farm_nsdi14, myth_vldb17, xenic_sosp21, ipipe_sigcomm19, linefs_sosp21} aimed at mitigating these issues, they have not fully overcome the inherent performance disadvantages of networking techniques.

The introduction of Compute Express Link (CXL) technologies~\cite{cxl-doc, cxl-shortdoc, cxl-paper}, particularly the CXL 3.0 specification, provides a promising solution for achieving both scalability and efficiency in transaction systems. This technology defines a Global Fabric Attached Memory (G-FAM) node, which allows compute nodes to access G-FAM with memory semantics, and to cache G-FAM data within their processor cache hierarchies. The CXL back-invalidation (BI) scheme maintains cross-node coherence by monitoring and invalidating outdated caches at the CXL fabric using a MESI-like protocol. With the optimized datapath, CXL achieves ultra-low link latency and sub-microsecond G-FAM access latencies~\cite{pond, directcxl, tpp_asplos23, cxl_anns_atc23, cxl_demystify, neomem}. Despite the potential, it remains unknown whether the coherence link of CXL suits transaction processing.

We demonstrate that the vanilla adoption of CXL-based memory sharing in transaction processing systems achieves performance far worse than expected, primarily due to two reasons. First, CXL incurs significant overheads due to "remote cache signal" process which is necessary for coherence maintenance. Here, data loads or stores on G-FAM may notify other nodes' caches to either invalidate or update their cachelines. This process involves multiple cross-node communication roundtrips over the CXL fabric, with latencies exceeding 800ns in our FPGA prototypes, rendering coherent access 1.85$\times$ slower than direct G-FAM DRAM access. Second, to maintain coherence, CXL requires the implementation of an inclusive Snoop Filter (SF) on the G-FAM fabric to monitor the states of cachelines at each node. The SF presents significant scalability challenges, as its centralized design limits the capacity for memory sharing and the number of compute nodes.

\hspace*{\fill}

\noindent \textbf{Insights. }
Our key insight is that the strict coherence model of CXL is overkill for a large fraction of memory accesses in transaction processing systems. In these systems, memory accesses can be classified into record accesses and metadata accesses, based on the type of data they engage. While metadata accesses require strict coherence to ensure synchronization correctness, record accesses typically adhere to well-established transactional consistency models~\cite{rss_sosp21, tcc_isca04, sitm_asplos14}, where a record’s store operations are finalized only at the point of transaction commitment. This setup allows for a degree of temporary incoherence among transactions that are still in progress, and transactions that are eventually aborted do not require maintained coherence. However, hardware cannot distinguish between these types without application-provided information. Previous research in concurrency control algorithms, such as SILO~\cite{silo_sosp13}, supports our observations. Nonetheless, without modifications to underlying hardware mechanisms, these approaches fail to resolve the SF scalability issues. Other works on transactional memory~\cite{logtm_hpca06, overlaytm_pact19, flextm_isca08, vtm_isca05} share similar concepts with our findings but are limited to a single socket and are not practical as they necessitate significant changes to processor architectures.

\noindent \textbf{\name-Primitive Innovations.}
To leverage this insight, we propose a \textit{hybrid} primitive architecture that maintains CXL’s strict coherence for metadata while decoupling coherence maintenance from the memory access process for record accesses. The \name~primitive transfers the decision of whether and when to achieve cache coherence from hardware to software, yet still utilizes CXL’s native hardware datapath for coherence operations. By strategically invoking the coherence API, specifically at the point of transaction commitment, transaction system developers can significantly reduce unnecessary cross-node cache coherence overheads. Additionally, software developers can further mitigate cache coherence latencies using techniques such as batching and co-routines, akin to traditional network-based systems.

\ifx\undefined\stale
\noindent \textbf{Our Approach. }
We propose~\name, the first CXL-native in-memory transaction processing system. We base~\name~on a key insight: \textit{The transaction software clearly knows when to synchronize modifications with other hosts better than the application-oblivious hardware coherence protocols.} 
As a consequence, we construct an unique loosely coherent memory layer on hardware, rather than the most processors (so as CXL) adopted strong coherence model. 
The memory layer follows a software-defined data coherency where the updates toward a record referring to the partial order specified by the transaction programs~\cite{TSTM}.
Transactions make in-situ reads and writes on records optimistically, and trigger the memory layer to publish their updates when commits. 
Specifically, we employ host's LLCs to buffer transaction's speculative updates, rather than allocating additional software buffers and do costly access redirection. 
By carefully choosing CXL protocols, we make the LLC incoherent on different hosts. 
When a transaction commits, the processor installs all updates to memory with explicit flushes. 
By doing so, \name~avoids unnecessary cross-host coherence messages associated with accessing records and saves the software buffers. 
Furthermore, we introduce a two-phase \textbf{hardware-assisted validation} to alleviate the heavy coherence messages associated with retrieving and validating per-record metadata~\cite{stm, nvm, distributed-tm} when validating conflicts. Specifically, \name~maintains a bloom filter for each on-the-fly transaction as an abstract of their read and write sets. 
A validation process first compares the transaction's bloom filter with other concurrent transactions. If the bloom filter indicates potential conflicts, it fall backs to standard per-record validation at the second phase. 
The key of high performance is to keep such bloom filters in hardware, and utilize CXL's cache coherence to reduce the latency of data movement between hosts. 
We evaluate \name~against state-of-the-art distributed transaction processing systems~\cite{farm, drtm} and the aforementioned baseline systems using various benchmarks~\cite{}. Our experiments demonstrate that \name~can achieve a throughput that is much higher than previous network-based designs. 
\fi



\noindent \textbf{Architectural Supports.}
We present holistic architectural support for our \name~primitive, designed with the following goals:
\noindent \textbf{G\#1: }
The architectural design should neither change host processor architectures nor CXL specifications.
\noindent \textbf{G\#2: }
The architectural design should ensure compatibility with various transaction processing systems' policies.
\noindent \textbf{G\#3: }
It is also important to limit implementation overheads to ensure the efficiency of the \name~primitive.



To meet these goals, we contribute a hardware-software co-design with three modules:
First, we introduce an \name~hardware agent (CTHW) at the G-FAM side. The CTHW modifies the semantics of host processors' loads and stores by remapping them to a node-private address space, and exposing cache synchronization with explicit side-path API calls.
Second, on the host software side, we design a lightweight \name~user-level library (CTLib) that allows applications to select primitives for every G-FAM allocation according to application-specific requirements.
Third, we propose a user-level runtime thread (CTRt) to monitor CTHW states and reconfigure it according to the use case to maintain efficiency.



\ifx\undefined\stale
We discuss how the \name~architecture meets the aforementioned design goals:
For \textbf{G\#1}, we limit our hardware changes to the device to be compatible with existing processor architectures and CXL protocols. 
The CTLib provides an allocation-specific primitive binding API, enabling current single-node systems to deploy \name~by only modifying memory allocation codes, meeting \textbf{G\#2}. 
To meet \textbf{G\#3}, we offload reconfiguration and interrupt handling to CTRt, keeping hardware logic deterministic and fast. Additionally, we keep CTLib off the datapath of memory loads and stores to avoid software overheads on common memory operations.
\fi
\noindent \textbf{Contributions. }
In summary, this work makes the following contributions: 

\begin{itemize}


    
    
    
    \item We discuss the network-induced performance degradation in traditional distributed transaction processing systems and illustrate how CXL-based memory sharing offers a viable solution (Sec.~\ref{sec:background}).

    \item We critically examine the limitations inherent in standard CXL primitives for memory sharing. Through quantitative analysis, we demonstrate that the performance is far below expectations, due to two key performance issues (Sec.~\ref{sec:motivation}).

    \item We introduce a comprehensive solution with a software and hardware co-design approach. We detail a suite of custom primitives (Sec.~\ref{sec:primitive}), and describe the systematic and architectural supports of \name~(Sec.~\ref{sec:details}). To the best of our knowledge, this is the first work studying CXL-based memory sharing on transaction processing systems.

    \item Based on the \name~architecture, we implement an exemplary key-value store with a broad exploration of design choices in transaction systems, based on the popular framework DBx1000~\cite{abyss_vldb14}. In evaluations using typical OLTP benchmarks, \name~reduces over 95\% of cross-node coherence traffic and achieves up to 2.08x throughput improvements.

\end{itemize}



\section{Backgrounds}   \label{sec:background}

\ifx\undefined\stale
\subsection{Distributed Transaction Processing}    \label{subsec:kvs_primary}




Transaction processing is the basic building block for modern datacenter applications, which provides a high-level abstraction that a single host executes transaction at a time and never fails. In current data center use cases, datasets are always partitioned into multiple shards, which are then distributed across multiple memory nodes. 

transactions typically span across nodes, so that they rely on networks to access records on remote nodes. 

\begin{figure}[t]
  \includegraphics[width=0.48\textwidth]{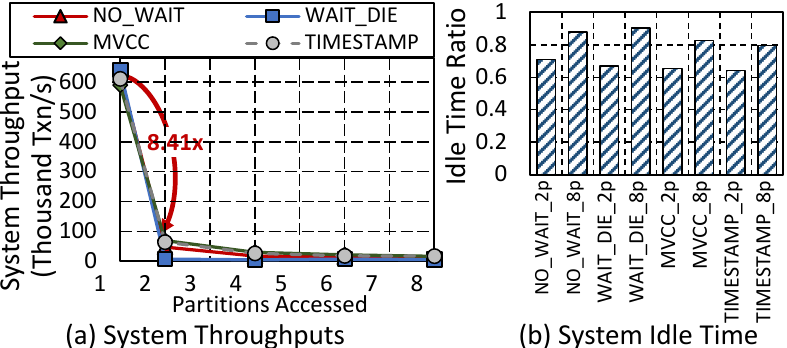}
  \caption{
  (a) varies the number of partitions that each transaction accesses and draws the system overall throughput. (b) breakdowns the time ratio of the processor being idle. 
   }
  \label{fig:sna-sda}
\end{figure}



However, cross-node networks may easily bottleneck the system performance, regardless of the software policies that are adopted. 
We test deneva~\cite{dbx1000_dist_vldb17, abyss_vldb14}, a popular distributed transaction processing framework, on 8$\times$r320 instances in CloudLab~\cite{cloudlab}. 
Figure~\ref{fig:sna-sda}~(a) shows the commitment throughput on various concurrency control policies. As we vary the number of nodes that each transaction accesses from one to two, all protocols degrades over 8.4x due to the network overheads. We further breakdown the processor execution time and observe over 60\% are spent on waiting for the network to complete, as shown in Figure~\ref{fig:sna-sda}~(b). This is a common case despite a variety of works optimize the latency of networks, such as remote direct memory access~\cite{ddio_ispass20, ddio-doc, farm_nsdi14, myth_vldb17, xenic_sosp21, ipipe_sigcomm19, linefs_sosp21} and smartnics~\cite{xenic_sosp21, clio_asplos22}. It's because the inherent overheads of PCIe copying, network card's asynchronous singling, and network stack processing~\cite{directcxl} required for every network transmission. 

\fi

\ifx\nocmt\undefined


\noindent \textbf{Strict Serializability. }

The aim of concurrency control is to provide a strict serializable consistency model of concurrent transactions. The model maintains an illusion of a single machine that executes transactions one at a time, in the order with respect to the real time~\cite{farm, drtm, compromise, timestone_asplos20, calvin_sigmod12, cicadia_sigmod17, tm_book, rss_sosp21, hekaton_sigmod13}. The key of the strict serializable consistency is the global order of the observed reads and writes~\cite{rss_sosp21, acid_79}. For example, consider a read in a KVS that returns the value written by a concurrent write, the read imposes a global constraint on future reads that they all must return the new value, even if the write has not yet finished. 
A typical transactional system guarantees this feature by the commit-abort mechanism, so that a transaction can commit if all of its operations obey this ordering constraint, or it should abort and revert the changes. 
This renders an important feature of transactional writes: a write within a transaction is non-atomic, so that it is issued during execution but can only be observed once the transaction commits. 
Figure~\ref{fig:bg-kvstore}~(b) illustrates an example, where transaction T1, T2, and T3 are trying to overwrite the value of A, and T4 is reading out A. 
One can see that the value of A changes at the point of transaction commits. 
T4 commits before T3, so that T4 can only read out the old value despite T4's read happens before T3's write in physical time. 

\fi

\subsection{Network-Based Transaction Processing}


Transaction processing is a fundamental component of modern datacenter applications, providing a high-level abstraction that simulates a single host executing a transaction at a time without failure~\cite{farm_nsdi14, compromise}. Current datasets exceed the memory capacity of a single node and are typically partitioned and distributed across multiple memory nodes. This setup necessitates cross-node networking communication for accessing remote records, which can involve great cost. Despite current systems~\cite{farm_nsdi14, myth_vldb17, xenic_sosp21, ipipe_sigcomm19, linefs_sosp21, grappa_atc15, guideline_atc16, fasst} employing various optimization techniques like co-routines and request batching to mitigate network overheads, their performance still greatly degrades if the workload inolves remote data accesses. We evaluate the state-of-the-art system DrTM-H under \rvs{an RDMA cluster with 8 CloudLab's x170 nodes~\cite{cloudlab}}. As the ratio of remote data access increases from 0\% to 100\%, its throughput degrades by over 53.3\%. 

We summarize two primary reasons for this degradation: First,\textbf{ network stacks' overhead}. Typical networking fabrics maintain a complex protocol stack to serve a wide range of hosts with robustness against packet loss, yet these stacks incur significant time costs. For example, \rvs{reading 64 bytes via RDMA can consume over 78\% of total latency in copying payloads between buffers within the network stack~\cite{directcxl}}. Second, \textbf{lack of cache coherence support}. Unlike single-node shared memory programming, networking primitives adopt a message-passing model that does not guarantee cross-node coherence at the primitive level. Consequently, it's common for distributed systems to avoid caching remote data~\cite{ddio_ispass20, ddio-doc, farm_nsdi14, myth_vldb17, xenic_sosp21, ipipe_sigcomm19, linefs_sosp21} to eliminate software overheads associated with managing coherence. Thus, these systems fail to exploit data locality in applications and involve network communication for every remote data access~\cite{fasst, learned_cache}.






\subsection{Compute Express Link Basics}


Compute Express Link (CXL) is an open industry interconnect standard based on PCIe 5.0 physical links, contributed by a variety of vendors from firmware~\cite{samsung, samsung_2}, operating systems~\cite{tpp_asplos23}, and datacenter providers~\cite{tpp_asplos23, pond}. The recent advancements in the CXL protocol (specification 3.0) introduce a hardware memory sharing programming paradigm~\cite{cxl-doc, cxl-shortdoc, cxl-paper}, which allows external memory, i.e., Global Fabric Attached Memory (G-FAM), to be disaggregated from processing nodes. Multiple compute nodes can share G-FAM by accessing it with standard memory semantics.


The CXL's memory accessing sub-protocol, \cxlmem, enables the read/write memory access primitive at a 64-byte cacheline granularity. Loads and stores via \cxlmem~are managed with the well-optimized hardware datapath in processors~\cite{quantitative_approach, book_cc}, enabling the access latency of G-FAM to achieve sub-microseconds, significantly lower than network-based approaches by an order of magnitude~\cite{legoos_atc19, gam, txcache_osdi10, learned_cache}. Another key improvement of CXL is hardware-enabled cache coherence. CXL 3.0 introduces a back-invalidation channel (\cxlbi) that allows endpoint devices to invalidate host CPU's caches. The \cxlbi~enables the endpoint to retrieve host's un-flushed dirty cachelines or invalidate stale cachelines using a MESI-style protocol.

\ifx\undefined\stale
CXL implements three sub-protocols between the Root Complex (RC) and the endpoints (EP). 
\cxlio~resembles traditional PCIe-like I/O primitives to be responsible for device controlling functionalities, such as device discovery, and interrupt handling. 
\cxlmem~enables the read/write memory primitive in a 64-byte cacheline granularity towards accessing device's attached memory. With \cxlmem, the host processor can treat device's internal memory as a host-managed device memory (HDM), which can be mapped to the physical address in the system memory just like local DRAM. 
\cxlmem~accesses are cachable and dirty cachelines are flushed back to the memory following standard cache eviction flow (typically writeback in modern x86 architectures~\cite{intel-doc, spandex_asplos18, book_cc}). 
\cxlcache~allows the EP to get processors' un-flushed dirty cachelines with a MESI style protocol. It enables EP-side compute units to invalidate or to fetch dirty cachelines from the host CPU's cache hierarchies, and allows the host CPU accesses device's caches in the same way. 
The memory-intensive accelerators like smartNICs~\cite{ccnic_asplos24} could benefit from it since they incorporate internal memory and private cache hierarchies, and potentially share data between the EP and host CPUs. 
\fi

\ifx\undefined\stale
Instead of reusing the host-device coherence sub-protocol \cxlcache, \cxlbi~is introduced to avoid the circular channel dependence on \cxlmem~that could potentially lead to deadlocks~\cite{cxl-shortdoc, cxl-doc, cxl-paper}. 
\fi




\begin{figure}[t]
  \includegraphics[width=0.47\textwidth]{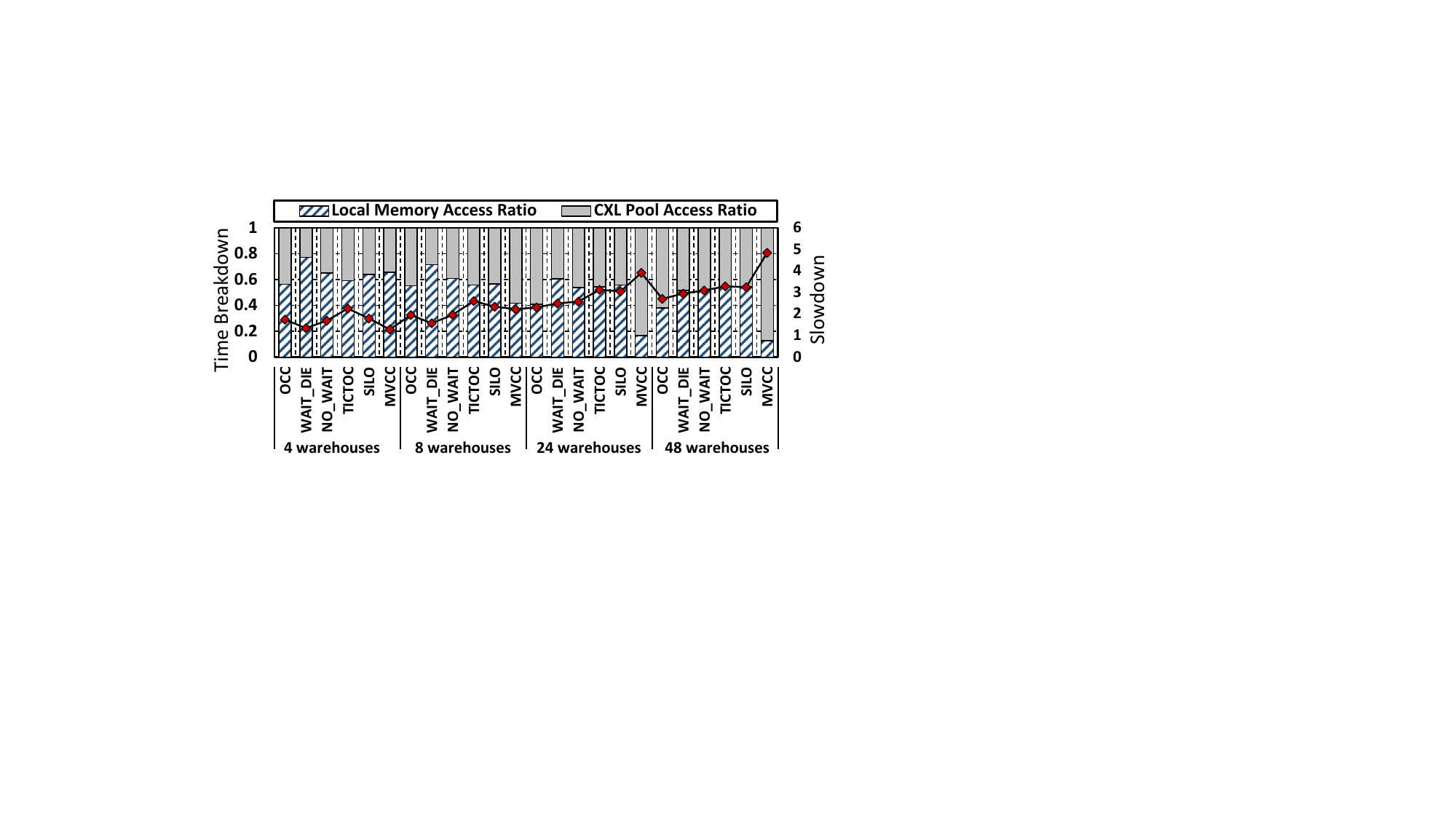}
  \caption{
  Relative slowdown than \textit{oracle}.}
  \label{fig:cxl_slowdown_breakdown}
\end{figure}

\subsection{CXL-Based Transaction Processing Systems}

One promising CXL-based approach for transaction processing rebuilds the remote procedure call (RPC) layer with a more efficient CXL implementation. This method adapts to the traditional networks' partitioned architecture, where each node exclusively owns a partition of Global Fabric Attached Memory (G-FAM) and uses RPCs to access partitions from other nodes. Hence, it enables existing network-based systems~\cite{hstore_damon16, citus_sigmod21, memsql_vldb16, voltdb, calvin_sigmod12} to migrate to CXL with almost zero code modifications. 
However, despite being carefully optimized, current SOTA CXL RPCs~\cite{hydrarpc_atc24, partial_sosp23} still suffer from performance overheads due to inherent software control overheads and cache flushing instructions. For example, the HydraRPC~\cite{hydrarpc_atc24} exhibits the averaged latency 3.22x longer than directly accessing G-FAM's DRAM, and consumes considerable CPU cycles on polling the RPC queue. Moreover, partitioned systems work great only if the workload is also perfectly partitionable. As the number of transaction-touched partitions increases, their performance degrade quickly~\cite{hekaton_sigmod13, silo_sosp13}. 
 

Another approach involves building a shared memory system leveraging CXL’s cache-coherent memory semantics. This approach employs single-node systems~\cite{silo_sosp13, abyss_vldb14, polyjuice_osdi21} that place shared data structures, including indexes, locks, and tuples, on the G-FAM, allowing worker nodes to run transactions in a multi-threaded manner. The \cxlbi~ensures hardware-level coherence across nodes. Compared to partitioned systems, this organization better aligns with CXL’s capabilities for cache coherence and exploits data locality, especially beneficial for indexes~\cite{masstree, learned_cache} and locking mechanisms. In this paper, we stick to this shared memory system where each transaction can access the entire dataset. Our evaluations in Section~\ref{sec:eval} shows that the \name-optimized shared memory system outperforms the CXL-RPC partitioned system at most 7.3x on poorly partitioned workloads.

\section{Motivations}   \label{sec:motivation}

\subsection{Scaling Transactions with Coherent CXL} \label{subsec:vanilla_primitive}

To investigate how current transaction systems perform under the CXL's shared memory, we deploy the popular transaction processing framework DBx1000~\cite{abyss_vldb14} on the G-FAM. We defer the evaluation details to Section~\ref{subsec:setup}. Figure~\ref{fig:cxl_slowdown_breakdown} illustrates the performance of six commonly adopted concurrency control algorithms on the TPC-C benchmark~\cite{tpc-c}. The numbers are normalized to an \textit{oracle} architecture that operates on local DRAM and consolidates all worker threads within a single socket. We group the time components based on the memory modules accessed by the transactions. It is evident that CXL memory sharing slows down all algorithms by 2.51x on average, with 46.89\% of the time spent on accessing G-FAM.
Algorithms that maintain multiple tuple versions, such as \textit{MVCC}, experience the most significant performance loss, since they require more CXL memory accesses to create new tuple versions and reclaim the stale ones than those maintaining only a single tuple version.

\ifx\undefined\stale
Due to the lack of CXL 3.0 IP, we enable memory sharing by hacking the IP's address translation to fold a half of physical addresses to map to the same DRAM address of another half. We implement an LRU SF and emulate the \cxlbi~with \cxlcache~since they have similar protocol constitutions. We leave the details to Sec.~\ref{subsec:setup}. We employ the popular transaction processing framework DBx1000~\cite{abyss_vldb14} and modify its memory allocation logic to locate the shared data structures such as indexes and tuples at the G-FAM, while preserve the thread-private data on the local DRAM. 
\fi

\subsection{CXL's Coherence Cost}  

CXL's memory sharing, along with most caching architectures~\cite{rvweak_pact17, armcm_cmb12, powercm_pldi11, rvweak_isca18, gam, txcache_osdi10}, adopts the \textit{strictly coherent} cache model~\cite{munin_ppopp90, rtm_isca14} to simplify software development. The basic design principle of this model is to ensure a single, global order for all memory locations. A memory store in the strictly coherent model is only completed once all peer nodes have observed it. This order is also known as the single-writer-multiple-reader invariant~\cite{book_cc}, which allows only a single node to execute a write primitive on a location or multiple nodes to execute read primitives at the same time.
In subsequent discussions, we refer to the primitives that adhere to this coherence model as ``CXL-vanilla.''

Maintaining such a coherence model requires complex hardware logic to send coherence requests and track the lists of sharing nodes. In CXL, this is achieved through an MESI-like ownership-based (write-back) coherence and directory-based communication~\cite{spandex_asplos18, book_cc, denovo_pact11, quantitative_approach}. 
Write-back coherence means that a cache write is not immediately reflected in the DRAM but instead marks the cacheline as dirty. Consequently, a cache read miss retrieves the content from a remote dirty cache rather than directly from the memory.
Directory-based architecture implies that the hardware maintains a directory to track which caches hold each cacheline and their states. A cache controller intending to issue a coherence request first sends it to the directory to determine which node is holding the peer caches. Such a directory is implemented as a Snoop Filter (SF) in CXL specifications.

\ifx\undefined\stale
A read miss starts with a memory read request from node-2 to the EP. 
When the EP receives the read request, it checks its internal directory and finds out node-1 is holding the copy. 
In case the node-1 has modified X, EP needs to send a back invalidation message via \cxlbi~to node-1 to get its X content. 
Node-1 thus responses its modified X to the EP, and the data is further forwarded to node-2, thereby resolving the coherence across node-1 and node-2. 
\fi

\subsubsection{Remote Cache Signaling Cost} \label{subsec:remote_fetching}

\begin{figure}[t]
  \centering 
  \includegraphics[width=0.47\textwidth]{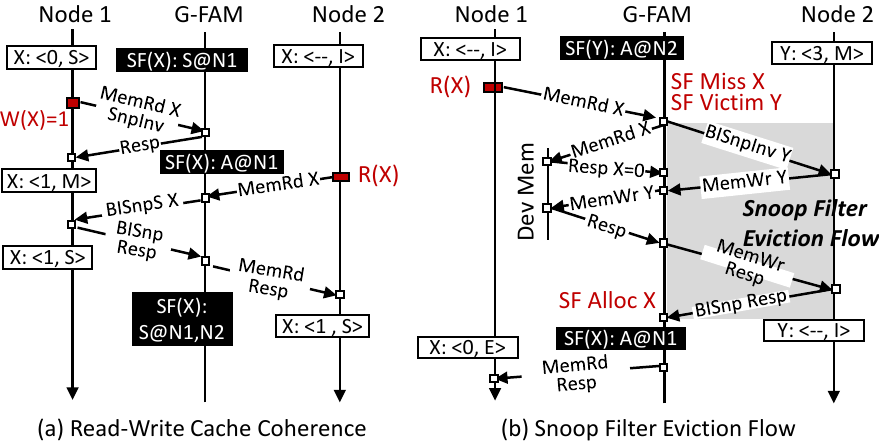}
  \caption{(a) Remote cache signaling to node-1 from a node-2's read. (b) SF eviction process incurred by node-1's read. }
  \label{fig:motivation-over-coherent}
\end{figure}

In order to maintain the strict coherence model, \vanilla~introduces a remote cache signaling process to invalidate a remote cacheline for exclusive ownership requirements or to retrieve the contents for a read miss. 
Figure~\ref{fig:motivation-over-coherent}~(a) illustrates an example where a cache read miss on node-2 retrieves a modified cacheline from node-1 via the CXL fabric.
This process involves two CXL request-response roundtrips between the host and the G-FAM node: one for node-2’s read request to the G-FAM, and another for the read request from G-FAM to node-1. Each CXL roundtrip is designed to take about 75ns with an ideal ASIC implementation~\cite{pond, tpp_asplos23, directcxl, cxl-shortdoc}, 
\rvs{
and the currently available FPGA-based platform exhibits an average latency of approximately 400ns~\cite{cxl-fpga-doc, cxl_demystify, hydrarpc_atc24}. We develop a real-world prototype system utilizing these off-the-shelf hardware. As illustrated in Table~\ref{tab:latency}, the remote cache signaling roundtrip takes approximately 850ns, which is 3.4 times greater than the latency of an ideal ASIC implementation. This gap is primarily due to the low operational frequency of FPGA's CXL IP and the intrinsic overhead associated with this FPGA's chiplet architecture. 
We will detail this prototype in Section~\ref{subsec:setup}.
}


\begin{table}
\caption{Roundtrip Latency Comparison}\vspace{-5pt}
\label{tab:latency} 
\resizebox{0.47\textwidth}{!}{%
\begin{tabular}{|c|c|c|c|c|}
\hline
             & Socket & NUMA  & CXL (Ideal) & CXL (Proto) \\ \hline
Core-to-Mem  & 92ns       & 145ns & 170ns                 & 456ns                 \\ \hline
Core-to-Core & 49ns       & 133ns & 246ns                 & 847ns                 \\ \hline
C2C/C2M    & 0.53x         & 0.91x    & 1.44x             & 1.85x                 \\ \hline
\end{tabular}%
}
\end{table}

\ifx\undefined\stale

\begin{figure}[t]
  \includegraphics[width=0.48\textwidth]{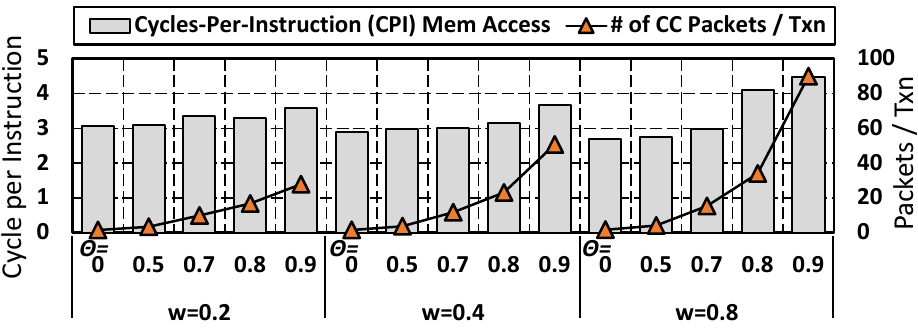}
  \caption{The stack indicates cycle-per-instruction on memory accessing. The line indicates the CXL packets per transaction. \red{G-FAM}}
  \label{fig:cc2read}
\end{figure}

To illustrate the performance effects of remote cache fetching, we employ a simpler benchmark YCSB~\cite{ycsb} that enables us to directly tune the read and write ratios. The memory accesses follow the zipf distribution controlled with a skew factor $theta$. A large $theta$ indicates a small portion of objects are being accessed by transactions, so that enlarges the cross-node memory sharing. Moreover, a larger write ratio increases the coherence traffic since a store would invalidate all peer caches. By increasing $theta$ from 0 (i.e. uniform accessing) to 0.9 (i.e. over 80\% accesses on 10\% tuples), we can see in Figure~\ref{fig:cc2read}, the time for an instruction to access memory increases 17\%, 27\%, and 66\% at 0.2, 0.4, 0.8 write ratios. 
We then observe the coherence packets received and sent by all nodes per transaction increases dramatically (55.9x) when the skew factor increases from 0 to 0.9 at the 0.8 write ratio. 
Even with fast IO connections, the remote cache signaling introduced by CXL-vanilla primitive still effects the system performance. 
\fi




In addition to the significant link latency, CXL introduces a unique coherence bottleneck when a read miss is served by the remote dirty cache. 
Previous works~\cite{ccnic_asplos24, pond} have adopted the non-uniform memory access (NUMA) architecture as a mimic of CXL memory. 
However, such a mimic is not adequate in the memory sharing since CXL exhibits relatively larger latency in remote cache signal time compared with conventional NUMA interconnects, such as the QuickPath Interconnect (QPI)~\cite{qpi-doc}. As shown in Table~\ref{tab:latency}, in most commercial NUMA architectures~\cite{directory_sp19, moesi_isca22}, reading a remote cacheline is not overly costly since the NUMA remote signal time is similar to accessing local memory. 
However, in CXL, the remote cache signal takes 44\% to 85\% more time than the latter, primarily because CXL requires two link roundtrips to inform other nodes, whereas NUMA only takes one. 
\rvs{
As the cluster size increases, the latency gap tends to expand, primarily due to the potential overhead introduced by retimers and CXL switches~\cite{pond} associated with the CXL link. An increase in roundtrip latency could exacerbate the overhead of remote cache signaling process.
}

\hspace*{\fill}

\noindent\fbox{%
  \parbox{0.47\textwidth}{%
     \textbf{Take-away\#1:} Remote cache signaling makes considerable performance overheads on G-FAM accesses. 
  }%
}

\ifx\nocmt\undefined
This renders an insight that storing frequently sharing objects in the CXL memory can serve reads and writes much faster than local caches, thereby we can benefit the performance by properly flushing the caches to DRAM. 
\fi









\subsubsection{Snoop Filter Scalability Issues}    \label{subsec:snoop_filter}

To enable cross-host cache coherence, the CXL endpoint should maintain a snoop filter (SF) to track the ownership of cachelines associated with the address space of G-FAM~\cite{directory_sp19, moesi_isca22, skylake_sf_doc}. The SF should be inclusive of the G-FAM's cachelines in all processors' local caches. This means the SF needs to track which nodes hold the cacheline and in what states, if the cacheline is located in at least one node's cache. Every unique G-FAM load or store should occupy an SF entry.






Compared with the SF in NUMA architectures, CXL SF faces unique scalability issues due to its centralized design. The CXL SF is located only at the G-FAM node and tracks all hosts' accesses, whereas NUMA architecture distributes SFs across every socket, with each SF tracking only remote accesses to its own socket.
We use the existing NUMA SF design as a mimic since a specific CXL SF design has not yet been proposed. To support a common rack-scale cluster with 16 nodes, the CXL endpoint is expected to maintain a 64K-set 11-way on-chip SF, which is 16x larger than a single NUMA SF in the current Intel Skylake architecture~\cite{directory_sp19, skylake_sf_doc}. In the worst case, where all processors' valid cachelines cache the G-FAM contents, the SF would require 1.8 billion entries and consume 14.4GB of memory. As the number of nodes increases, the CXL SF should expand correspondingly and could easily become a system bottleneck.





\ifx\undefined\stale
We track the 

In a general rack-scale cluster with 16 nodes, and a 11-way 4K-set on-chip SF (respecting with the NUMA SF in Skylake), the SF could only support 2.75MB caches in total, which is only 0.1\% of the processors' internal caches.
Despite we can adopt DRAM as the backup of the SF to enlarge its available capacity, tracking the worst-case usage of on-chip caches are still inevitable. For example, tracking the 16 node cluster with 32 core in each node requires 1.8 billion SF entries, which will take 14.4 GB memory. 
\fi

Conflicts on the SF lead to an SF eviction flow that back-invalidates other cachelines~\cite{cxl-doc, cxl-shortdoc} to make room for SF allocation. Figure~\ref{fig:motivation-over-coherent}~(b) illustrates an example: a read on X, which is not currently recorded by the SF, incurs an SF eviction on Y since they map to the same SF slot. The EP must wait for the eviction flow to reach node-2 before responding to node-1's read request. Such a process involves three CXL roundtrips on the critical path, which take over 400ns more than the remote signaling process.

\noindent\fbox{%
  \parbox{0.47\textwidth}{%
     \textbf{{Take-away\#2: }}  The centralized snoop filter causes scalability issues of CXL-vanilla primitives.
  }%
}

\subsection{Do transactions require strict coherence ?}       \label{subsec:overkill}

We observe that the coherence model adopted by the \vanilla~primitives is overkill for most transaction processing systems. We use the key-value store (KVS) as an example. A typical KVS comprises two data fields: the \textit{record field}, which stores the actual data (KVS's value) managed by the system, and the \textit{metadata field}, which contains the system's internal data structures to ensure transaction correctness, such as locks, timestamps, indexes, and latches.



\begin{figure}[t]
  \includegraphics[width=0.45\textwidth]{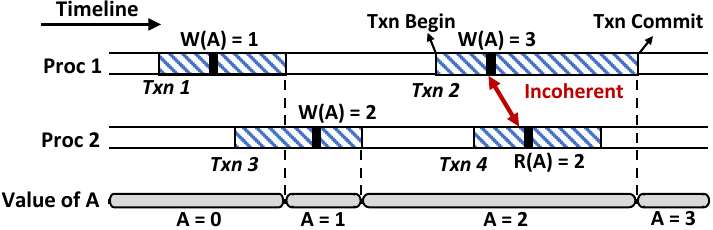}
  \caption{The strict serializable consistency. }
  \label{fig:txn_consistency}
\end{figure}

Accesses to the \textit{record field} follow a much looser order than what CXL's strict coherence model provides.
Specifically, accesses to the \textit{record field} are all wrapped within transactions that are committed as all or none. These accesses thus adhere to specific transactional consistency models~\cite{rss_sosp21}, such as strict serializability~\cite{tcc_isca04} or snapshot isolation~\cite{sitm_asplos14}. These consistency models permit reads from two or more transactions to observe temporarily incoherent values. Figure~\ref{fig:txn_consistency} illustrates an example of a strict serializable consistency model, where the read from Txn-4 precedes Txn-2's write, despite this read occurring after the write in physical time.

Such counter-intuitive ordering is maintained by the concurrency control algorithms in transaction processing systems.
Therefore, there is no need for the hardware to maintain the strict cache coherence model, which incurs substantial performance overheads (Take-away \#1 and \#2). As an alternative, \textbf{the hardware could simply maintain the partial order required by transactional consistency, allowing data incoherence according to the transaction's execution status and synchronizing data only at the commitments of transactions~\cite{munin_ppopp90, tcc_isca04, vtm_isca05, threadmarks_tc94, rtm_isca07, overlaytm_pact19, lamport_tc79, newdefinition_isca98}.}

Prior works on single-node concurrency controls, such as SILO~\cite{silo_sosp13}, align with our findings but their solutions are orthogonal to us. Specifically, SILO aims to avoid unnecessary coherence traffic incurred by \textit{metadata} writes that are required for read-only records' accesses, but it can not mitigate the remote signal costs associated with \textit{record} accesses themselves, hence it can not address the remote singling cost on records accesses. 
Moreover, SILO-like concurrency controls only modify the software algorithms, which do not address SF issues without changes to the underlying hardware coherence mechanisms.

\hspace*{\fill}

\noindent\fbox{%
  \parbox{0.45\textwidth}{%
     \textbf{{Take-away\#3:}} Transaction processing systems allow the record field accesses to be temporal in-coherent. 
  }%
}

\begin{figure*}[t]
\begin{minipage}{0.33\textwidth}
\resizebox{\textwidth}{!}{%
\begin{tabular}{|l|l|}
\hline
Proc. P1 @N1 & Proc. P2 @N2 \\
\hline
$I_1$: St/L-St $\alpha$ 1 & $I_4$: St/L-St  $\beta$ 1 \\
$I_2$: r1 = Ld/L-Ld $\alpha$ & $I_5$: r3 = Ld/L-Ld $\beta$ \\
$I_3$:r2=Ld/L-Ld($\beta$+r1-1) & $I_6$:r4=Ld/L-Ld($\alpha$+r3-1)\\
\hline
\multicolumn{2}{|l|}{\begin{tabular}[c]{@{}l@{}} CXL-vanilla forbids but \name~guarantees: \\  r1 = 1, r2 = 0, r3 = 1, r4 = 0 \end{tabular}}    \\
\hline
\end{tabular}%
}
\caption{L-Ld and L-St are node-private}
\label{fig:litmus_1}
\end{minipage} %
\hfill
\centering
\begin{minipage}{0.33\textwidth}
\resizebox{\textwidth}{!}{%
\begin{tabular}{|l|l|}
\hline
Proc. P1 @N1 & Proc. P2 @N2 \\
\hline
$I_1$: St/L-St $\alpha$ 1 & $I_5$: St/L-St $\beta$ 1 \\
$I_2$: r1 = Ld/L-Ld $\alpha$ & $I_6$: r3 = Ld/L-Ld $\beta$ \\
\textbf{$I_3$: GSync $\alpha$}    & \textbf{$I_7$: GSync $\beta$} \\
$I_4$:r2=Ld/L-Ld($\beta$+r1-1) & $I_8$:r4=Ld/L-Ld($\alpha$+r3-1) \\
\hline
\multicolumn{2}{|l|}{\begin{tabular}[c]{@{}l@{}} \name~forbids: r1 = 1, r2 = 0, r3 = 1, r4 = 0 \end{tabular}}    \\
\hline
\end{tabular}%
}
\caption{GSync broadcasts stores}
\label{fig:litmus_2}
\end{minipage} %
\hfill
\begin{minipage}{0.32\textwidth}
\resizebox{\textwidth}{!}{%
\begin{tabular}{|l|l|}
\hline
Proc. P1 @N1 & Proc. P2 @N2 \\
\hline
$I_1$: St/L-St $\alpha$ 1 & $I_5$: St/L-St $\beta$ 1 \\
$I_2$: r1 = Ld/L-Ld $\alpha$ & $I_6$: r3 = Ld/L-Ld $\beta$ \\
\textbf{$I_3$: Wd $\alpha$}    & \textbf{$I_7$: GSync $\beta$} \\
$I_4$:r2=Ld/L-Ld($\beta$+r1-1) & $I_8$:r4=Ld/L-Ld($\alpha$+r3-1) \\
\hline
\multicolumn{2}{|l|}{\begin{tabular}[c]{@{}l@{}} \name~forbids: r3 = 1, r4 = 1 \end{tabular}}    \\
\hline
\end{tabular}%
}
\caption{Wd withdraws stores. }
\label{fig:litmus_3}
\end{minipage} %
\end{figure*}

\section{\name~Overview}    \label{sec:primitive}

\ifx\stale\undefined
\subsection{The Over-Coherent Problem}
In a well-established transaction processing system, operations toward the \textit{records} are already well ordered referring to a \textit{strict serializable consistency} by concurrency control algorithms. 
These operations form only a small subset of possible interleaving orders.
To be specific, transaction processing exhibits an optimistic nature on coherence~\cite{farm_nsdi14, drtm, compromise, timestone_asplos20, calvin_sigmod12, cicadia_sigmod17, tm_book, rss_sosp21, hekaton_sigmod13} so that it's legal for multiple transactions to hold the same record simultaneously in either an unchanged or a speculatively modified form. The different versions are serialized at the boundary of transactions, but no ordering limits between uncommitted operations. 
The ordering limits are further relaxed when one conflicted operation is from an aborted transaction, since it's unnecessary to serialize a memory operation that will be discarded. 
It provides an opportunity to delay the coherence point only at the commit phase, and helps to reduce the coherence cost. 

However, 
the CXL-based memory sharing, as well as most MESI-based caching architectures~\cite{rvweak_pact17, armcm_cmb12, powercm_pldi11, rvweak_isca18, gam, txcache_osdi10}, adopt a \textit{strict coherence} model~\cite{munin_ppopp90, rtm_isca14}. This model ensures accessing to every memory location following the single-writer-multiple-reader (SWMR) invariant, which enables only a single core that may write to the location or multiple cores that may read the location. 
It's a pessimistic ordering model that addresses data races at every point they happened, by only allowing one version at a time. 
All aforementioned performance drawbacks are the cost to maintain the SWMR invariant: the remote fetching is essential to address potential data races on ownership upgrades (Shared to Exclusive) and silent writes (Exclusive to Modified); the SF is essential in the directory-based protocol to route coherence requests to proper places. 




From the software perspective, a transaction should prevent its uncommitted write from being observed by others. But the cache coherence conducts the broadcasts constantly. 
To this end, a transaction system achieves the write isolation via pessimistic locks~\cite{s2pl_csur81, abyss_vldb14}, or optimistic write buffers~\cite{silo_sosp13, hekaton_sigmod13, occ_tbs81, farm_nsdi14}. There's a wide debate on which approach performs better, and some works~\cite{tictoc, polyjuice_osdi21, cormcc_atc18} dynamically select both according to the application contention scenarios. However, few discussion is made on altering the underlying cache coherence model to make it fit for transaction processing, since it's overly ambitious to change cache designs and integrates extensive logic into processor circuits. Fortunately, it's possible in CXL-based memory sharing since the CXL specification only specifies the coherence protocol but leaves the freedom of cache design to developers. 
\fi

\ifx\stale\undefined
\subsection{Optimistic Cache Coherence Protocol}


\name~leverages the loose coherence nature to eliminate the high coherence cost in CXL. \name~proposes a release consistency (RC)~\cite{rc} like memory consistency model that delays the version serialization to when the transaction commits. 

RC is a widely adopted relaxed consistency model~\cite{threadmark, munin, txcache}. It distinguishes between two types of memory operations: ordinary and synchronization. RC introduces two synchronization operations: acquire and release. The acquire operation is performed before a process enters a critical section, ensuring that any updates made in other critical sections by other processes are visible. The release operation is performed when exiting a critical section, signaling that all updates made within the critical section are now visible to other processes. 
RC follows two key ordering principles: 
First, operations within a critical section can be reordered, but synchronization operations (acquire and release) enforce a partial ordering. Second, operations before a release cannot be reordered to after it, and operations after an acquire cannot be reordered to before it.

This brings up the following three benefits: 

Challenges: 
Commercial cluster deployment
Supporting variety of concurrency control algorithms
Supporting essential components of IMDB

Changing the cache design to fit transaction processing is overly ambitious since it integrates extensive logic into processor circuits and requires non-travial efforts for verification. Fortunately, it's possible in CXL-based memory sharing since the CXL specification only specifies the coherence protocol but leaves the freedom of cache design to developers. 
\fi

\subsection{\name~Primitives}

To exploit the incoherence opportunity in the record field (\textbf{Take-away\#3}), we propose a hybrid primitive design that implements different protocols depending on the field accessed by the primitive. Users can employ the strictly coherent protocol for the \textit{metadata field} to ensure the correctness of atomic operations. For the \textit{record field}, \name~advocates for a new, loosely coherent protocol. In the following discussion, we use ``\name~primitive'' to indicate the \textit{record field} primitives for simplicity. The \name~primitive decouples costly cross-node coherence operations in CXL from normal memory accesses. It comprises four fundamental memory operations: Local-Load (L-Ld), Local-Store (L-St), Global Synchronization (GSync), and Withdraw (Wd).

L-Ld and L-St retrieve and put data from/to the shared memory, and their effects are limited to the requesting node, making no coherence demands. 
We utilize litmus tests~\cite{rvweak_pact17} to compare L-Ld and L-St with standard CXL Loads (Ld) and Stores (St). 
As depicted in Figure~\ref{fig:litmus_1}, P1 forwards the value from I1 to I2 and I3 locally without making I1 visible to P2 at N2, allowing r2 and r4 to both be 0 simultaneously. In contrast, with \vanilla~primitives, if both I2 and I5 return the value 1, it implies that stores I1 and I4 must have been globally observed, hence r2 and r4 cannot both be 0.

The GSync broadcasts a node-private value globally, while the Wd withdraws updates made by preceding L-Sts that have not yet been broadcast. 
The synchronization primitives should be specified for a particular address.
As depicted in Figure~\ref{fig:litmus_2}, following the GSync primitives, both P1 and P2 can observe the new value stored by I1 and I5. Consequently, I4 and I8 cannot both be 0, similar to the behavior observed with \vanilla~primitives.
As illustrated in Figure~\ref{fig:litmus_3}, Wd cancels the value stored by I2 so that it is invisible to I8.

\ifx\undefined\stale
\noindent \textbf{Benefits for the primitive innovation. }
The GSync and Wd grant the software control to selectively advertise a memory access. As we will discuss in later, L-Ld and L-St would not incur any coherence overheads on execution. By carefully invoking the GSync primitives referring to the system-adopted transaction consistency model (Sec.~\ref{subsec:implementation}), for example, only synchronizing writes at commits when using strict serializability model, the system developers could avoid unnecessary remote cache signaling penalty from those uncommitted or aborted operations. 
Moreover, the GSync and Wd primitive removes the SF checking from the critical path of G-FAM accessing. It thus tolerates higher latency that could be exploited to improve the spatial efficiency of SF. 
\fi



\ifx\define\stale
We define the coherence ordering rules resembling the definition of weakly memory consistency~\cite{newdefinition_isca98}: 
\begin{enumerate}
    \item Primitives are strongly coherent within a node so that processors at the same node observe only one primitive order. If primitives $p_1$ and $p_2$ belong to the same node, we say $p_1$ \textit{happens before} $p_2$ if $p_1$ precedes $p_2$ in the node's order. 
    
    \item \textit{Syncs} are strongly coherent across nodes so that processors at any nodes observe only one \textit{Sync} order. We say a the $Sync_1$ \textit{synchronizes before} $Sync_2$ if $Sync_1$ precedes $Sync_2$ in this order.
    
    \item If primitives $p_1$ and $p_2$ belongs to different nodes, we say $p_1$ \textit{happens before} $p_2$ iff. $p_1$ \textit{happens before} a $sync_1$ and a $sync_2$ \textit{happens before} $p_2$ and $sync_1$ \textit{synchronizes before} $sync_2$. 
    
    \item If 
\end{enumerate}
\fi




\ifx\define\stale
In a transaction processing system, operations on the \textit{record field} are already well ordered referring to a \textit{strict serializable consistency} or more loose ordring models such as snapshot isolation by the software runtimes. 
These operations form a limited subset of possible interleaving orders than general parallel codes, thus it's no need to keep the strict pessimistic coherence as general cases. 
To be specific, operations would follow the two key principles on data coherency. 
First, the \textit{record} reads and writes within a transaction should be coherent, so that a read reflects its preceding writes on the same data object. 
Second, the \textit{record} reads and writes observed by different transactions could be incoherent if one belongs to an aborted or an uncommitted transaction. On the other words, data coherency is only preserved between committed operations.

These rules indicate an optimistic coherence limit on the \textit{record field} so that it's legal for multiple transactions to hold the same record simultaneously in either an unchanged or a speculatively modified form, and these forms are only need to be serialized at the commit phases. On the other hand, the \textit{metadata field} requires a strict coherence guarantee that a write must grant the inclusive ownership of 
the target address and all nodes should observe the same global value order. 

To achieve the both worlds, we propose a heterogeneous coherency primitive that distinguishes memory operations by addresses (Sec.~\ref{sec:model}).
We adopt the strict coherence model for the \textit{metadata field}. But for the \textit{record field}, we propose a loosely coherent primitive that decouples the main memory read/write with remote cache invalidation and data fetching. Memory operations would hit the node private memory space that introduces no coherence issues, and data coherence is delayed to the transaction commitment phase. The decoupled optimistic coherence resolves aforementioned CXL's performance drawbacks in two aspects. 
First, reducing remote fetching cost traffic on the only true data races. 
Second, we remove the remote fetching from the performance-critical reads. So that we adopt a tracking data structure replacing SF to track the ownership of each node. It trades a false-positive rate by significantly reducing the SF's memory demands. 
\fi

\ifx\undefined\stale
\noindent \textbf{G\#1: Hardware compatibility.} 
The design should not disturb neither host processor architectures nor CXL specifications.
Because recent processor has plenty of confidential and undocumented features~\cite{tsx, directory_sp19}. Alterations to them could introduce unforeseen effects in performance or even violate system correctness, posing significant challenges in system verification and debugging.

\noindent \textbf{G\#2: Universal support on transaction systems. } 
The design should ensure primitive compatibility with existing transaction processing systems. These systems are different in consistency models, data structure organization, and scheduling policies, and asks for the \name~system to provide a single-node like memory management methods, such as \textit{malloc} and \textit{free}. 
Keeping compatibility with the various software systems enables our proposal to be treated as an inexpensive feature that can be adopted without effecting other services. 

\noindent \textbf{G\#3: Low performance overhead. } 
It's also important for the design to introduce minimal overhead on the critical path, ensuring that the proposed primitive has performance advantages to CXL-vanilla primitives. It requires \name~implements memory loads and stores with the minimized hardware overheads. No software overheads, such as interrupts or context switching, are allowed since a primitive only takes hundreds of nanoseconds, but the software intervention will introduce several microseconds overheads. 
\fi


\begin{figure}[t]
  \centering
   \includegraphics[width=0.45\textwidth]{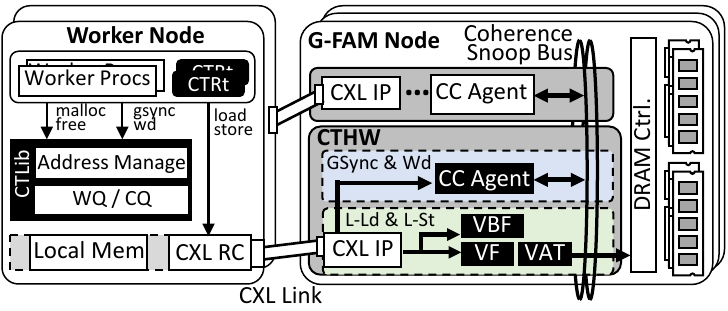}
  \caption{
    \name~overview.
  }
  \label{fig:overview}
\end{figure}

\subsection{Design Challenges}

\noindent \textbf{Challenge\#1: Being Compatible with Processor Cache Design.} 
Implementing L-Ld and L-St requires isolating accesses across nodes. 
Merely disabling the \cxlbi~for shared memory accesses does not suffice to implement L-Ld and L-St due to cache eviction issues. Recall that CXL uses write-back protocols~\cite{cxl-shortdoc, cxl-doc}, meaning that stores always hit local caches. However, cache eviction can inadvertently leak modified contents from the private cache to the shared memory, allowing subsequent loads from other nodes to observe these modified contents. 
Worse still, cache eviction is managed by uncore cache agents, which operate transparently to cores.

\noindent \textbf{Challenge\#2: Being Compatible with CXL Protocols.} The CXL protocol does not differentiate between cacheline flushing and eviction; both are treated as \cxlmem~writeback requests. This conflation prevents the use of cacheline flushing as a synchronization point (to implement GSync), which was possible in earlier weakly consistent architectures~\cite{rvweak_pact17, rvweak_isca18, spandex_asplos18, lrcgpu_micro16}. Moreover, current x86 architectures do not support the Wd semantic that merely self-invalidates a cacheline without writing it back.

\noindent \textbf{Challenge\#3: Being Compatible with Software Policies.} To fully exploit the advantages of the hybrid primitive, our system should allow application processes to choose which primitive to use for each data structure they allocate. Simply partitioning the address space to map to different primitives is too rigid to accommodate the diverse requirements of various use cases.



\ifx\undefined\stale
Simply discarding the hardware cache coherence for record field may work well at functionality. Such a strawman approach disables the back-invalidation channel by default and a transaction explicitly flushes what it writes when commitment. 
Once received the cache write back requests, the home node (CXL device in this example) invalidates all peer caches to force processors reading the up-to-date data from memory. 
This approach is widely adopted in weakly consistent memory, such as GPUs, IBM Power~\cite{rvweak_pact17, rvweak_isca18, spandex_asplos18, lrcgpu_micro16}. 
Synchronization is maintained by the coherent metadata field which supports the atomic operations. 

However, this strawman approach has two performance drawbacks. 
First, dirty writes could be flushed back to memory at any time before the transaction commits, due to either cacheline eviction or compiler-added memory fences. 
Without modifying the CXL protocol, the device is unable to distinguish the cacheline eviction from cacheline flush, since they share the same cacheline write-back flow in \cxlmem~\cite{intel_cxl_pub, cxl-paper, cxl-doc}. 
As a consequence, this approach incurs the unnecessary invalidation snooping request at every dirty cacheline eviction. It not only wastes CXL bandwidth, but also enlarges the average latency for an LLC cache miss.
Second, the explicit cache flushing instruction (e.g. CLFLUSH in Intel~\cite{intel-doc} processors) is costly in modern processors, since they need to clear its write buffers to keep memory consistency. As pointed by previous work~\cite{dynamicamo_isca23}, the cacheline flush takes xx more time than a memory write. 
Moreover, this approach still requires the centralized SF to route invalidation snooping requests correctly.

However, the cache eviction would break such isolation since it would leak the modified cachelines from the private cache to the shared memory. If a read from other nodes is issued right after the eviction of a written cacheline, it would observe the written data thereby violating primitive semantics. The cache eviction could happen at any time due to cache conflicts. In typical x86 architectures, the cache eviction is managed by the uncore cache agents which are transparent to the processor's pipeline. Hence it's impossible to address the eviction issue by inserting special instructions such as CLFLUSH or FENCE to codes, neither by changing the processor architectures (\textbf{P1}). 

\fi

\ifx\stale\undefined
\name~connects each hosts with PCIe Gen5 x16 links to provide similar memory bandwidth of local DRAM. The connection uses either on-board connections or the xxx links extended from the PCIe slot on mother board. Ideally, each host directly connects to each \name~with individual physical links. This results in a cluster with N nodes and M devices using NM links, and occupying M PCIe x16 slots on each processor. It's challenging for the commercial system to connect a large amount of \name~cards due to limited PCIe slots on the IO die. 
Users can adopt CXL switch to reduce the occupation of IO slots by connecting processors to the switch's upstream ports, and connecting devices to the downstream ports. This approach will lead to a higher latency by the estimated 100ns for traversing a CXL switch. 
We limit the memory sharing field within a rack with up to 16 nodes since this scale keeps the sweet point between performance and actual use cases. 
\name~assigns an individual CXL driver IP to each host despite they may share the same physical link when adopting CXL switch. Each \name~device manages up to 1TB~\cite{anns, cxl_samsung} DRAM with device internal memory controllers. 
\fi

\ifx\stale\undefined
Commercial cluster deployment
Supporting variety of concurrency control algorithms
Supporting essential components of IMDB
\fi


\ifx\stale\undefined

\subsection{The Metadata Field: Atomic Offloading}

\noindent \textbf{The write amplification problem. }

\noindent \textbf{Our approach. }

\fi

\ifx\stale\undefined

In such a model, \textbf{every write appears atomically} to the memory subsystem, in order to keep the single-writer-multiple-reader invariants for every object. A read must returns with the value of most recent write to that object. 
This feature is held in both hardware cache architecture such as SC, TSO, RMO, Alpha, and ARMv8, and in most software distributed shared caches~\cite{gam, }. CXL achieves this via a Global-Observation (GO) based MESI protocol, as illustrated in Figure~\ref{fig:motivation-over-coherent}~(a). 

However, transactions preserves a more loose \textit{strict serializable consistency model}. \blue{briefly introduce other models, i.e. read-isolation} This model maintains an illusion of a single machine that executes transactions one at a time, in the order with respect to the real time~\cite{farm, drtm, compromise, timestone_asplos20, calvin_sigmod12, cicadia_sigmod17, tm_book, rss_sosp21, hekaton_sigmod13}. Record operations are packed and ordered at the granularity of a transaction. The key of the strict serializable consistency is that the global order of the observed reads and writes~\cite{rss_sosp21, acid_79} should refer to the serialized time of their belonging transactions, but not the physical time they are executed. 
We refer this phenomenon to the \textbf{non-atomic write} in transaction processing, where a write is logically executed at some point of the transaction, but it takes effect only after the transaction commits. 
A read is consequently able to return a stale value if the write is serialized after that read. 

\fi

\ifx\nocmt\undefined
Considering a read from transaction A returns the value written by a concurrent write from transaction B, the read imposes a global constraint on all A's reads that they all must return the new value, even if the write has not yet finished. 
\fi




\subsection{Architecture Overviews}

%

\name~integrates multiple components to address these challenges. 
As Figure~\ref{fig:overview} shows, we employ a hardware agent (CTHW), which implements \name~primitives in the G-FAM to address the cache eviction problem for L-Ld and L-St (\textbf{Challenge\#1 and \#2}) and to enable synchronization primitives with a different mechanism (Sec.~\ref{subsec:vms},~\ref{subsec:vat}, and~\ref{subsec:vsf}). 
Additionally, a user-level library (CTLib) facilitates the selection of memory primitives for individual data structures (\textbf{Challenge\#3}), enhancing software integration with the hardware capabilities (Sec.~\ref{subsec:system_integration} and~\ref{subsec:memory_mapping}). 
Moreover, a helper thread (CTRt) adjusts the hardware configuration to maintain system efficiency (Sec.~\ref{subsec:vat_resizing}). 

\name~adopts a non-transparent interface for memory management calls such as \textit{cxl\_alloc} and \textit{cxl\_free}, but keeps accessing approaches the same as local memory. System developers can bind the primitive to each allocation with passed arguments and load or store data as they would with local memory. At the point of transaction commits, the developer should invoke the GSync primitive to synchronize record writes globally, and if the transaction aborts, they should call the Wd primitive to withdraw stores.



\ifx\undefined\stale
\noindent \textbf{Hardware Agent (CTHW). } The CTHW is a device-only hardware module that implements \name~primitives in ASIC by design. To avoid cache eviction from leaking the modified contents (\textbf{Challenge\#1}), CTHW adopts a view memory shim (Sec.~\ref{subsec:vms}) model to implement the copy-on-write semantics, and introduces the architectural supports (Sec.~\ref{subsec:vat}) that achieves both high efficiency and low hardware overhead. To overcome the limits of write-back caches and CXL protocols (\textbf{Challenge\#2}), CTHW introduces a queue-based approach (Sec.~\ref{subsec:vsf}) for the GSync and Wd primitives.

\noindent \textbf{User-Level Library (CTLib). } CTLib proposes a memory-mapped primitive choosing approach (Sec.~\ref{subsec:memory_mapping}) to allow the fine-grained primitive binding on data structures. Moreover, CTLib is carefully integrated into OS (Sec.~\ref{subsubsec:system_integration}) to avoid software overhands on memory loads and stores (\textbf{Challenge\#3}).



\noindent \textbf{Helper Thread (CTRt). }The performance of CTHW data path is highly effected by the application memory accessing characteristics, thus the CTRt monitors the hardware status and perform CTHW reconfiguration (Sec.~\ref{subsec:vat_resizing}) to avoid CTHW bottlenecks the system (\textbf{Challenge\#3}). 

\fi

\ifx\stale\undefined

The conflicts between CXL's atomic writes and transaction's non-atomic writes requires an overlayed memory layer to isolate uncommitted writes from being observed by others. It's typically provided by concurrency control, and is termed as ``version management''. 
Commonly adopted approaches include in-situ updates and write buffers. In in-situ updates, transactions directly update the tuples and record the values that have been overwritten. The recorded values are recovered in case of aborts~\cite{s2pl_csur81, abyss_vldb14}. This approach assumes the pessimistic execution and depends on the control fields, as we will discuss in Sec~\ref{}, to mutually exclusive. 
On the other hand, the write buffer approach allocates temporal local buffers for each transaction, and redirects speculative writes to the buffers while keeps the shared tuple unchanged. It flushes the buffer contents when commits~\cite{silo_sosp13, hekaton_sigmod13, occ_tbs81, farm_nsdi14}. 
Modern concurrency controls interleave both of them within a transaction execution according to the application contention status~\cite{tictoc, polyjuice_osdi21, cormcc_atc18}, or allow multiple record versions co-exist~\cite{mvcc, mvcc_vldb17}. 

The version management is the well-known bottleneck, especially for the optimistic algorithms such as \textit{OCC}, \textit{TICTOC}, and \textit{SILO}~\cite{full-story, abyss, deneva}. The main overheads come from buffer allocation and reclamation and memory copying for each written tuple. Figure~\ref{fig:cxl_version_management} shows the version management bounded algorithms spend over half of the time on this procedure, which become the main reason of the up to 60\% overall performance loss. 

\fi

\ifx\stale\undefined

The cache copy is flushed to the CXL DRAM only when a transaction commits, and is discarded once aborts. 
To this end, we eliminate the need of software memory overlay and its related performance cost. 
We leverage existing CXL subprotocol requests to implement the decoupled store without disturbing host processors' architectures. 
However, many issues remain to ensure corretness, we leave the detailed discussion in Sec.~\ref{subsec:}.

Different with all of the overlayed approaches, we resort to directly change the CXL hardware consistency model to make it match transaction processing. We 
To be precise, we decouple a store instruction to two memory references: store issue, and store commit. 
Stores that are issued but not committed are kept within the node's caches for the duration of the transaction in order to keep invisible from other nodes. 
No conventional, CXL's MESI style cache coherence protocol is used to maintain a global order of conflicted writes. As a consequence, it is legal for multiple nodes to hold the same record with conflicted modified forms. 
The cache copy is flushed to the CXL DRAM only when a transaction commits, and is discarded once aborts. 
To this end, we eliminate the need of software memory overlay and its related performance cost. 
We leverage existing CXL subprotocol requests to implement the decoupled store without disturbing host processors' architectures. 
However, many issues remain to ensure corretness, we leave the detailed discussion in Sec.~\ref{subsec:}. 

\fi

\begin{figure}[t]
  \centering
  \includegraphics[width=0.47\textwidth]{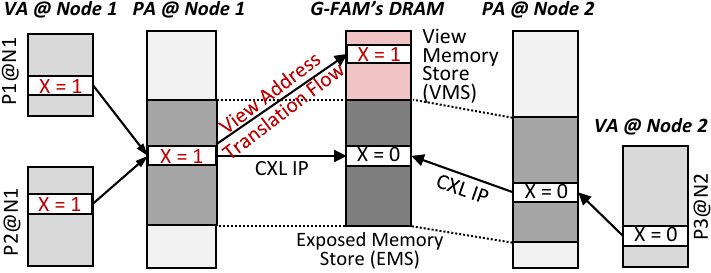}
  \caption{View memory shim. }
  \label{fig:view_model}
\end{figure}

\section{\name~Architectural Supports}     \label{sec:details}

\subsection{View Memory Shim}   \label{subsec:vms}




In the following discussion, we first focus on the architectural support of the loosely coherent model and leave the integration of hybrid models to Sec.~\ref{subsec:memory_mapping}. 
In this section, we illustrate the core conceptual model of our design, named view shim layer. The view shim layer operates between the node's physical address space and the G-FAM DRAM address space, aiming to redirect memory accesses from nodes to the appropriate G-FAM DRAM locations according to L-Ld and L-St semantics. We adopt the concept of ``view'' from relational databases~\cite{mvc_micro23, pageoverlay_isca15, overlaytm_pact19}. 
As Figure~\ref{fig:view_model} shows, each physical address can be mapped to a DRAM location called a view shim, in addition to a regular DRAM location mapped by the CXL IP. At a high level, when a physical address has both an original DRAM mapping and a view shim mapping, the L-Ld and L-Sts are served by the view shim. Only addresses not present in the view shim are accessed from the regular DRAM location. 
Each node can own at most one view shim for a shared object, but views from different nodes are isolated. In this example, the physical object X is mapped to both the original DRAM address space and a view shim in node-1, while node-2 only has the DRAM mapping. Consequently, the process in node-1 observes X=1, while the process in node-2 observes X=0.

Logically, a view shim is created on a memory store and is merged with the original memory location upon a GSync call, resembling copy-on-write semantics. 
Physically, the view has two statuses: on-chip and overflowed. The on-chip view resides within the processor's cache hierarchy as a common dirty cacheline and transitions to the overflowed status once it is evicted from the processor. A view starts in the on-chip status since any store would first load the cacheline to the processor followed by a store hit. 

The on-chip view naturally fits the L-Ld and L-St semantics as it is only observed by local loads and stores. However, the G-FAM needs to ensure that overflowed views are not observed by other nodes. To achieve this, we allocate a DRAM space called the view memory store (VMS) to buffer the overflowed views, acting as an evicted cache. We term the DRAM space that stores original shared contents as the exposed memory store (EMS) and maintain the address mapping between EMS and VMS via the node-wise view address translation flow.

\ifx\undefined\stale
\noindent \textbf{Benefits for the view memory shim model. }
First, by redirecting written back caches at the EP side, we successfully keep the L-Ld and L-St private to node while being transparent to both processor architecture at hardware and the OS's virtual memory subsystems. 
Note that this approach is not allowed in traditional SMP architectures but is only enabled by the CXL, since CXL HDM allows hardware changes on the memory manager~\cite{neomem}. 
Second, The host cache is the major store of view shims, which are fast and well verified, but the device's VMS is only accessed when a memory access misses the entire processor's cache hierarchy. This brings two performance benefits. (1) The traffic related with view shim management is greatly reduced. (2) The amount of view shims that VMS manages is typically small. 
\fi

\ifx\stale\undefined
Keeping the loose coherence with read and write redirection allows three distinct advantages.
First, L-LD and L-St avoid incuring the remote fetching process, thus much faster than the CXL-vanilla loads and stores. 
Second, our approach is transparent to the virtual memory system of current OS by laying under the physical address. The current address translation component such as TLB and page tables would not observe the view shim (\textbf{P2}). 
Third, as we will describe the next section, our design fully exploit the write back nature of the typical x86 processors' cache designs (\textbf{P1}). The host cache is the major store of view shims, which are fast and well verified, but the device's view memory store only acts as the victim cache that is accessed when a memory access misses the entire processor's cache hierarchy. This incurs two performance benefits. (1) The traffic related with view shim management is greatly reduced. (2) The amount of view shims that view memory store manages is reduced. 
\fi






\ifx\stale\undefined
While the view abstraction imposes simple memory access primitives, there are several key challenges to efficiently implement the proposed primitives. 
First, the hardware should check if a cacheline belongs to the view shim. 
Second, the hardware should ensure the consistency of committed writes. 
\fi

\subsection{View Address Translation Flow}      \label{subsec:vat}

\begin{figure}[t]
  \centering
  \includegraphics[width=0.45\textwidth]{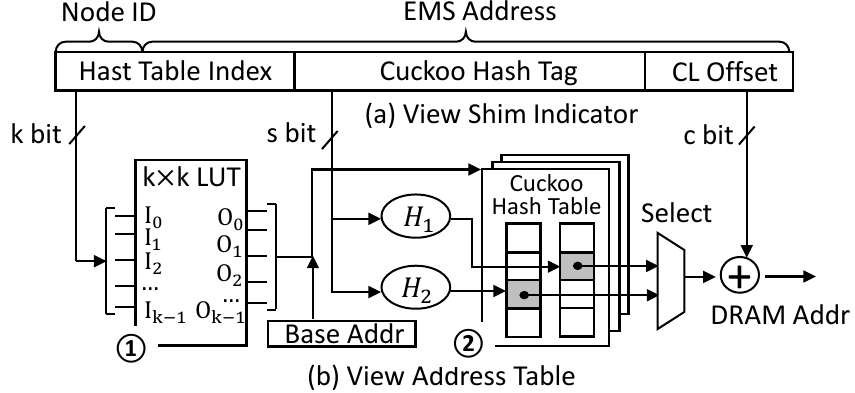}
  \caption{(a) The format of view shim indicator and (b) View address table (VAT) components. }
  \label{fig:vms}
\end{figure}


The view address translation flow controls whether the G-FAM returns a read request with the view in VMS or the original content in EMS, and it allocates space for overflowed views. It does not change the CXL IP but operates behind it. The CXL IP retrieves and translates the physical address to the EMS address~\cite{cxl-fpga-doc, cxl-doc}. 
Our flow takes this decoded information as input and outputs the target DRAM address according to the view shim model. We architecturally define three data structures: the view address table (VAT), which manages the mapping from the view shim indicator to the VMS address; the VMS filter (VF), which precedes the VAT to sift its query traffic; and the VMS Back Filter (VBF), which tracks the processor caches.

\subsubsection{VAT Details. }

A VAT entry (VATE) holds the address mapping at cacheline granularity. The VAT resembles OS page tables in functionality, but traditional radix-tree-style page tables are not suitable for our scenario since address translation involves page table walks that potentially traverse all levels of the tree sequentially~\cite{cuckoo_pagetable_asplos20, clio_asplos22}. To address this, we adopt a flattened cuckoo hash table architecture that bounds the number of view address translations within a limited number of DRAM accesses. 
As Figure~\ref{fig:vms}~(b) shows, we organize the VAT as a variable list of cuckoo hash tables~\cite{cuckoo_hash, cuckoo_pagetable_asplos20}. 
The rationale behind choosing cuckoo hashing is to trade the low cost of L-Ld with the insertion overhead of L-St, 
since an L-Ld typically lies on the critical path, whereas an L-St does not, due to current processors' out-of-order execution and asynchronous cache eviction~\cite{book_cc, rvweak_isca18, rvweak_pact17, quantitative_approach}.

\ifx\undefined\stale
The cuckoo hash table maps one key to two possible hashing locations with different hash functions. The element is stored in at most one of these locations at a time, but it can move between its hashing locations. An L-Ld incurs a cuckoo hash query, which checks both hashing locations and succeeds if it is found in either of them. If no valid VATE is found, it returns with the original EMS address.
An L-St incurs a cuckoo hash insertion, which places an element in one of its two possible entries. If the selected entry is occupied, the algorithm kicks out the current occupant, and re-inserts to the other hashing location. The process keeps until no occupant is evicted, or meets the number of maximum tries. 
\fi

An on-chip lookup table (\oone) stores the hash table base addresses, and the hash tables are located in the VMS (\ttwo). We use the first k bits to calculate the table's base address and the following s bits to query the cuckoo hash tables. 
\rvs{
The cuckoo hash table maps one key to two possible hashing locations using different hash functions. An element is stored in only one of these locations at a time but can be relocated between them. An L-Ld operation introduces a cuckoo hash query, which checks both locations and returns success if the element is found in either. If no valid VATE is located, it reverts to the original EMS address.
An L-St operation initiates a cuckoo hash insertion, placing an element in one of the two possible entries. If the chosen entry is already occupied, the algorithm displaces the current occupant, attempting to re-insert it at the alternative hashing location. This displacement process continues until it succeeds without eviction or reaches the maximum number of allowed retries.


There are two potential performance inefficiencies in this VAT design. First, a VAT insertion may require memory allocation for the cacheline content, in addition to inserting the VATE to the hash table. To mitigate this, we pre-allocate buffers for cacheline contents associated with the hash table. Second, cuckoo hash insertion may fail after reaching the maximum number of retries. Upon such a failure, the CTHW blocks all accessing to the hash map, sets the resizing error bit, signals the CTRt to resize VAT and retries the insertion (Sec.\ref{subsec:vat_resizing}).
Nonetheless, the occurrence of such an insertion failure is very infrequent in practice, as the CTRt actively oversees VAT occupancy to keep it within a moderate level. A threshold of 0.6 guarantees that an insertion is typically accomplished within 6 retries. We will discuss this at Sec.~\ref{subsec:latency_sweep}. 
}







\begin{figure}[t]
  \centering
  \includegraphics[width=0.49\textwidth]{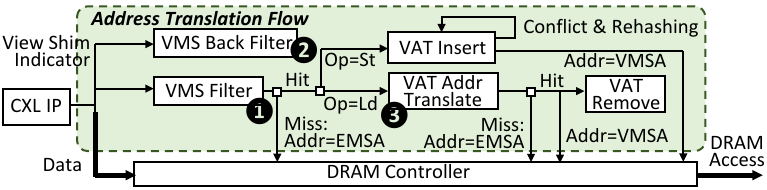}
  \caption{Address translation process for L-Ld and L-St. }
  \label{fig:pipeline}
\end{figure}


\subsubsection{VF and VBF Details. }

Despite the VAT optimizing DRAM address translation, it still impacts performance since every memory access must check the VATE in DRAM. However, most VAT queries would miss for two reasons. First, stores are relatively rare in real-world transaction applications~\cite{snow_osdi16, port_osdi20}, so the VMS is typically sparse. Second, most views reside in host processor caches and are seldom written back to the VMS due to widely adopted LRU-based cache eviction policies.

To leverage this feature, we introduce an on-chip VMS Filter (\onee~in Figure~\ref{fig:pipeline}, VF for short) to sift out VAT queries. A read on the shared memory would query the VF first, accessing the VAT only if the filter outputs a positive hit. Otherwise, the read falls back to the original EMS address. 
CTHW also tracks the host processor's caches to invalidate stale cachelines at synchronization (detailed in Sec.~\ref{subsec:vsf}). Consequently, each CTHW maintains a VMS Back Filter (\twoo, VBF for short) that works in reverse to the VF: a DRAM load inserts into the VBF, and a dirty write-back removes from the VBF. 
Compared with the original CXL SF, the VBF has three advantages:
First, VBF is organized in a distributed way, so that each VBF only tracks the local processor's cache, avoiding a single VBF becoming the system's bottleneck. 
Second, VBF queries and inserts are parallel with VF accesses, thus introducing minimal overhead to the memory access critical path. 
Third, the VBF exhibits much higher spatial efficiency when implemented with approximate data structures, such as elastic counting bloom filters~\cite{elasticbf_tc22}.

\ifx\stale\undefined

\subsection{Memory Address Management}

Similar to conventional main memory mapping, the EMS's addresses are mapped one-to-one to the physical memory address space of each node based on a node-wise address bias.
The CXL IP maintains such bias at the device enumeration phase (Section.~\ref{Sec.}). 
When a host accesses HDM, it passes the physical address to EP via the CXL link, then CXL IP at EP side subtracts the physical address with the corresponding node's base address and gets the EMS address. 
The VMS stores two things: view shim contents in the overflowed status for every node, and a view address translation table (VAT) that indexes the view shim contents. 

As Figure~\ref{fig:vms}~(a) illustrates, each object in the VMS is specified with an indicator that is a bit vector concatenated by the view owner's node id and the EMS address of the original object. VAT resembles the current operation systems' page table that maps the view indicator to the DRAM address. However, the traditional radix-tree-style page table does not fit our scenario, since looking up a translation involves a page table walk potentially all the levels of the tree in a sequential way. It incurs a pointer-chasing process wiht multiple DRAM accesses. 



We propose a flattened VAT that bounds the address translation to a fixed two DRAM accesses (\textbf{Principle 3}). Unlike normal virtual memory system that keeps address mapping at 4KB pages or larger, each view address table entry (VATE) manages the address mapping for a cacheline to align with the granularity of CXL requests. VATEs are allocated with the granularity of a large chunk. 
Accordingly, the view shim content's indicator is split into three segments, as Figure~\ref{fig:vms}~(b) shows. The most significant 12 bits are used to calculate the chunk base address, and we adopt the full sized on-chip lookup table to keep such a mapping. The following 43-6 bits are used to calculate the VATE chunk bias which are managed by a two-way cuckoo hash. An address translation checks the LUT and hashes for the chunk bias in parallel, the results are joint to determine the VATE address. The translation process could fail due to either the chunk does not exist, or the cuckoo hash can not find a valid VATE, it indicates the PA has no view mapping, and thus the process returns the EMS address.


The key insight of this cuckoo hash table is to trade the low look-up latency with the cost of insertion, since queries are introduced by a processor's read, which typically lies on the critical path, but an insertion is issued by a dirty cacheline eviction, whose latency does not effect the performance much due to existing cache optimization methods such as write buffer and non-blocking cache design. 
As Figure~\ref{fig:vms}~(b) shows, the cuckoo hash table maps one key to multiple possible hashing locations with different hash functions. The element is stored in at most one of these locations at a time, but it can move between its hashing locations. A look-up in cuckoo hashing checks all the possible hashing locations of an element, and succeeds if it is found in either of them. Insertion in cuckoo hashing places an element in one of its two possible entries. If the selected entry is occupied, the algorithm kicks out the current occupant, and re-inserts to the other hashing location. The process keeps until no occupant is evicted, or meets an number of maximum tries.

A well-known problem with the cuckoo hashing is the insertion failure. As occupancy increases, the insertion attempts could be large. Hence, it requires the VAT resizing, which requires costly and complex logic. As a consequence, we move it to the software runtime to keep hardware clean (Sec.~\ref{sec:implementation}). 
Fortunately, the VAT rehashing is quite rare because the view occupancy is typically low due to two reasons. First, a cacheline is only inserted to the view when it is both dirty and evicted from the entire host's cache hierarchy. Second, a view content is removed when a HDM read hits the view, or the writer transaction ends (Sec.~\ref{subsec:cold_path}). As the evaluation in the real-world memory traces (Sec.~\ref{}), the number of chunks is always under xxx.


As we discussed in Sec.~\ref{}, the VMS hash map keeps the overflowed dirty views, so that each dirty cacheline written back would issue the VMS insertion. 
According to the view primitive, any memory reads should check the VMS to see if there exists a view mapping, in addition to directly returns with the EMS content. 
Querying the VMS could be expensive when hash conflicts occur, hence we architecturally define a view filter (VF) that stays before every VMS hash map to sift the traffic. The VF is an on-chip bloom filter that tracks the content of each VMS.
A read should query the VF first, if it returns the positive, then the read could query the VMS to get the view. Otherwise, the read request falls back to the normal EMS access. The VF may incur false-positive error that reports positive for a non-exist entry, but it will not result in correctness problem. If the VF hit but the VMS misses, the read still falls bath EMS accesses. 

The usage of hash map incurs the challenge of hash collisions that could overflow a bucket. A typical hash-based data structure relies on conflict chaining or open addressing to solve the overflows, but both of them require the query to make unbounded number of DRAM accesses or costly pointer chasing. In order to bound VMS query to a constant number (typically one) of DRAM accesses, we adopt the multi-copy hashing to resolve the collision at the insertion time. 

\noindent \textbf{VMS Management (Principle 2)}. 

including failed insertion and hash resizing. When an insertion cannot find the proper victim after a specific number of tries, 


and 2) the view table (VT). Each view table entry (VTE) holds the mapping between a physical address from a specific node to an address of view shim frame. 
Similar to the traditional virtual memory abstraction, each host is given an impression that is working on a private, continuous section of HDM memory. But physically, the host-managed HDM could be redirected to a view shim, or be changed by other hosts at some data-race free points (e.g. transaction commitment). 
In common memory accesses, the VMS is invisible to both applications' processes and operating systems. The creation, querying, and deletion of VTE is automatically managed by the \name~hardware, as we detailed in the next section.

\noindent \textbf{Memory access flow. } 
The memory access flow serves memory reads and writes with a private view for each node. \name~adopts a copy-on-write resource management approach by architecturally defining two data structures: view address transactor (VAT) and the state tracker (ST). 
Resembling the page tables in OS, VAT keeps the address mapping between the virtual view address space and the physical DRAM address space. 
VAT only tracks the cachelines that are written back dirtily to reduce the space cost. 
To be specific, when a RC incurs a dirty writeback, the VAT will redirect this write to the node's view space (Sec.~\cite{detail}) by allocating a free slot. 
Then it records the redirected address of the slot to serve future reads. Consistently, when a RC issues a cacheline read, the VAT will check whether any previous writeback on the same cacheline exists. If so, the VAT redirects the read to view space, otherwise the VAT returns with the shared object of the consistent snapshot. 

The VAT querying could be expensive when the VAT is large, hence we introduce the ST to keep an abstract to reduce VAT visits. 
The ST tracks two datasets: the evicted dirty cachelines kept in VAT and the cachelines resides at the host caches. We base ST on the customized bloom filters (Sec.~\ref{detail}), so that it trades the high query performance for potential false-positive errors. A miss in the ST indicates a cacheline not exist in the queried status set, and a hit indicates that the cacheline probably exists. We introduce a dynamic sizing approach to keep the false-positive ratio within an acceptable boundary. 
\fi

\begin{figure}[t]
  \includegraphics[width=0.47\textwidth]{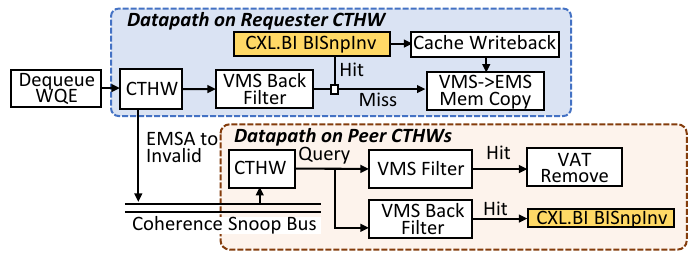}
  \caption{Synchronization datapath for GSync and Wd.}
  \label{fig:cmt}
\end{figure}

\subsection{View Synchronization Flow}  \label{subsec:vsf}




\subsubsection{Task Offloading Scheme. }

Different with the view address translation flow that works ``on the path'' of memory accesses, the GSync and Wd primitives are implemented in task offloading scheme. 
The endpoint maintains a queue pair for each host: a work queue and a completion queue. The host posts the addresses for GSync and Wd in the work queue element and polls the completion queue to synchronize for completion. The work queue is a circular buffer, while the completion queue is a bit vector where each bit maps to a work queue element.  
Both queues are located in the shared memory between hosts and devices and are accessed coherently utilizing the \cxlcache~sub-protocol. 
Inspired by the CC-NIC~\cite{ccnic_asplos24}, we apply an inline signal scheme to ensure only one coherent read is needed to retrieve the entire work queue element. Instead of signaling work queue changes with an explicit flag, we embed a valid bit into each work queue element. The endpoint polls the header element's valid bit and reads the payload once it observes that the valid bit is set. To further reduce coherence traffic, we limit the size of completion queue to a cacheline to enable one \cxlcache~read to get all pending completion queue elements. 

%

%



\subsubsection{Achieving Coherence Across Nodes. }

Synchronizing a dirty view involves two phases: invalidating the cachelines on remote nodes and merging the dirty view with the EMS.  
As Figure~\ref{fig:cmt} shows, we refer to the CTHW executing the GSync as the \textit{requester} and other CTHWs as the \textit{peers}. 
During remote cache invalidation, the \textit{requester} broadcasts an invalidation message on the coherence snoop bus. 
Each \textit{peer} checks its own VF and VBF to determine if it holds the corresponding cacheline in either on-chip or overflowed status. If the view resides in the VMS, the \textit{peer} simply invalidates the VAT entry to remove it. If it resides on-chip, the \textit{peer} invalidates the host processor's cacheline by issuing a standard \cxlbi~invalidation request~\cite{cxl-doc, cxl-shortdoc}. The CTHW ignores the written-back value since it's outdated. 

The content merging phase encounters two scenarios: the modified cacheline either remains in the host's cache or has already overflowed to the VMS. The \textit{requester} checks its VF and VBF to determine the status. If the cacheline is on-chip, the \textit{requester} sends a \cxlbi~invalidation request to the host to force a cacheline eviction, then writes the evicted data to the EMS. Otherwise, the \textit{requester} performs a memory copy from the VMS to the EMS and invalidates the VATE.

\ifx\stale\undefined
\noindent \textbf{Discussion and Optimization. }
The synchronization flow works on a bus-based snooping protocol, which may incur scalability issues due to cores scrambling for the bus. However, this would rarely bottleneck our design due to two reasons. 
First, the bus only transmits the addresses to invalidate but avoid cacheline data. This approach greatly reduces the bus bandwidth requirements. 
Second, the invalidation requests are only introduced by the software-introduced GSync primitives, and avoids the unnecessary coherence traffics incurred by pessimistic coherence model (discussed in Sec.~\ref{subsec:overkill}). 
For a the larger cluster that achieves the higher system throughput, we could further improve the bus throughput by limiting the coherence scale, adopting multiple buses that serve different address fields, but we leave it to future works. 
\fi

\ifx\stale\undefined

To address aforementioned problems, we propose \name, a hardware-software co-designed memory shim layer between CXL memory and user level services, as Figure~\ref{fig:overview} shows.
\name~takes care of the coherence issue as a replacement of hardware but preserving hardware-comparable performance. 
\name~reduces the costly cross-node coherence and SF evictions. 
\name~introduces no modifications to neither processor architectures nor CXL protocols to be practical, and it also keeps compatibility with common in-memory database services such as indexing, logging, and garbage collection. 

\fi

\begin{figure}[t]
  \centering
   \includegraphics[width=0.43\textwidth]{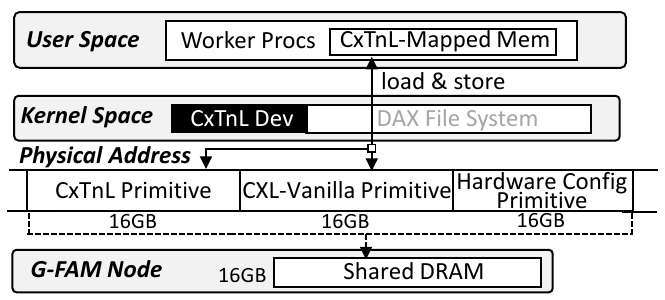}
  \caption{The \name's memory mapping to OS. }
  \label{fig:runtime} 
\end{figure}

\ifx\stale\undefined
This work focuses on the transactional distributed key-value stores that partition the records with primary key, as Figure~\ref{fig:bg-kvstore} shows.
The KVS preserves the strict serializable consistency for user transactions. 
This work optimizes for the online transaction processing (OLTP) applications with the main concern on optimizing system throughput as well as processing latency. In practice, OLTP transactions are (1) short-lived, (2) access a small number of records at a time, and (3) are repeatedly executed with different input parameters~\cite{dbx1000_dist_vldb17}. 

CXL-based memory pool extends the shared-address system beyond a physical node. 
Despite the lack of practice, it is easy to facilitate a single-node concurrency control to execute on a cluster by allocating the tuples on the CXL shared memory. 
Nevertheless, simply deploying existing approaches is not yet an appropriate option due to CXL's high coherence latency and the drawbacks of centralized SF architecture. 

Specifically, during the PCIe enumeration, the host driver would query the size of EP's internal memory and allocate the physical address fragment for it. The base address of such fragment is then told to the EP via the \cxlio's base address register (BAR) mechanism. 

\fi

\subsection{Operating System Integration}       \label{subsec:system_integration}

\ifx\undefined\stale
The CXL-backed memory can be exposed as a NUMA memory node or as a direct-accessed file. Carefully selecting the memory mode is crucial for \name's functionality. Pond~\cite{pond}, TPP~\cite{tpp_asplos23}, and the Linux kernel expose the HDM as a zero-core virtual NUMA (zNUMA) node, i.e., a node with memory but no cores. This approach manages memory with existing NUMA-aware memory management provided by OS (numactl in Linux). It imposes no need to change the application codes, however, only able to control the memory allocation at the application granularity. The current OS only support specifying the NUMA balancing strategies, but can not control which memory node that a specific data structure locates on. 
\fi



Resembling the Persistent Memory Development Toolkit (PMDK)~\cite{pmdk}, \name~adopts the Direct Access (DAX) mechanism to manage the shared memory~\cite{directcxl, cxl_anns_atc23, partial_sosp23}. We expose the device as a DAX file and allow the application to \textit{mmap} the G-FAM to its virtual address space. We do not adopt previous approaches of exposing the G-FAM as a zero-core virtual NUMA node~\cite{pond, tpp_asplos23}, i.e., a node with memory but no cores, since it is costly to use current NUMA control approaches to precisely specify the memory node at allocation. 

In this work, we set all G-FAM pages as ``reserved'' in the BIOS to prevent OS intervention, such as page swapping. This ensures that RTLib has full control to allocate the entire G-FAM to a contiguous virtual address segment for applications. 
The combination of DAX mode and page reservation ensures effective address translation among the virtual address space that CTLib operates on, the physical address space that the on-chip cache operates on, and the endpoint's DRAM address space that CTHW operates on, as it maintains a linear address mapping between these spaces.

\ifx\undefined\stale
The DAX mode takes full control of each allocation to the proper memory node. During the PCIe enumeration, the \name~driver queries the size of EP's internal memory and allocate a continuous physical address fragment to map it. The base address of such fragment is then told to the EP via the \cxlio's base address register (BAR) that resembles PCIe devices. 
We set all HDM pages as ``reserved'' at BIOS in order to prevent OS intervention such as page swapping out. To this end, \name~enables user-level runtime to take full control of the memory management by not changing the OS. 
The \name~library allocates entire HDM to a continuous virtual address segment at the application process initialization. The library addresses the shared memory objects with the bias and the base address pointer of the segment. 
\fi


\subsection{Memory-Mapped Primitive Selection}  \label{subsec:memory_mapping}



To enable multi-primitive selection, we expand the concept of memory-mapped I/O. 
As Figure~\ref{fig:runtime} illustrates, the \name~driver exposes the G-FAM to a physical address region with three times larger capacity. Each segment maps to a different primitive, including the CXL-vanilla primitive with hardware-managed strict coherence, the \name~primitive with decoupled coherence, and 
\rvs{
a special hardware configuration primitive for internal data structures of \name, such as VAT's hash tables, VMS contents, and work/complete queues. 
Applications could specify the CXL-vanilla primitive or the \name~primitive via CTLib's APIs, while the hardware configuration segment is reserved for \name~internal usage so that it is transparent to users. 
To avoid overlap between the application data and \name~data, we adopt a virtual address management strategy where the application address segment grows from the bottom base address upward, while the hardware configuration segment expands from the top address downward. When two segments overlap, i.e. no unused memory exists, an out-of-memory error occurs. 
}
\rvs{

We allow transaction processing systems to manage application address segments with customized allocators but allocation should be aligned with cacheline to match \cxl~protocol granularity. Conversely, CTRts are required to manage the hardware configuration segments. CTRt handles memory allocation and reclamation in fixed-sized chunks. Previously freed chunks are stored in a doubly linked list. Allocating a chunk in the hardware configuration segment first searches this list using a first-fit strategy. If no adequate chunks are available, CTRt expands the hardware configuration segment to satisfy this allocation. While advanced memory management strategies such as segregated free lists could enhance performance, delving into these methods is beyond the scope of this paper, thus we defer their exploration to future works.


}

\ifx\stale\undefined
The primitive selection is done at data allocations. 
CTLib provides memory allocation and reclamation APIs, i.e. \textit{cxl\_alloc} and \textit{cxl\_free}, being similar to traditional multi-threaded programming but require the primitive argument that following operations will obey. 
When the EP receives the HDM access, it chooses the primitive implementation based on which segment the physical address belongs to. 
In order to avoid the complex ordering issues across primitives, the CTLib allows each virtual address is bounded at most one primitive at a time for a process. 
By design, the CTLib on different nodes should communicate with each other to avoid conflict on hardware configuration space allocation. Our current implementation statically partition the segments for each node since the configuration space is typically small. CTLib do not address conflicts on application data structure allocation since it should be harmonized by the applications. 
\fi

\subsection{VAT Resizing} \label{subsec:vat_resizing}

The VAT resizing mechanism adapts the concept of page swapping deamon of virtual memory subsystems. To be specific, we introduce a helper thread CTRt to monitor the VAT's occupancy and conducting VAT resizing. At every $I$ milliseconds, the CTRt checks the occupation ratio of each hash table. The ratio is managed by the CTHW at hardware and is exposed through the hardware configuration address space. We introduce VAT expansion in this section, and VAT shrinking works in the reverse manner. If a table $t_{old}$ meets the occupancy threshold $r$, the runtime initiates a VAT expansion procedure to increase the $t_{old}$ to a table with $k\times$ equivalent capacity. The CTRt increases the number of hash tables to accommodate $t_{old}$'s VATEs. CTRt creates $k-1$ additional cuckoo hash tables with the same hash functions. CTRt scans the table base address LUT and moves the $t_{old}$'s VATEs with the prefix $s_i$ to the new hash table $t_{new_{i\%k}}$. The table bias of the VATE in $t_{new_{i\%k}}$ is the same to the $t_{old}$ thus avoiding the rehashing cost. After migration, CTRt updates the on-chip LUT to point the prefix to newly created hash tables. If the LUT is full that each $s_i$ points to an unique table, CTRt falls back to the rehashing. 

\rvs{To prevent data races between CTRt and worker threads, CTRt temporarily blocks any access to the hash table that is manipulated by the VAT resizing process. This involves negligible impact on overall performance, as VAT resizing occurs infrequently. The reasons are twofold. First, the VAT occupancy capacity is determined by the number of \textit{pending writes} from uncommitted transactions. Given that OLTP transactions are typically read-intensive and short-lived~\cite{snow_osdi16, port_osdi20}, the volume of pending writes is inherently limited in OLTP workloads. Second, the large on-chip caches is the primary location of pending writes, leaving VAT to manage only a minor fraction of the pending write set.
}

\section{Implementation and Evaluation}    \label{sec:eval}

\subsection{Exampled Transaction Key-Value Store} \label{subsec:implementation}

We construct a shared record key-value store (KVS) based on the \name~architecture. Similar to recent works~\cite{mtcp_nsdi14, xenic_sosp21, ipipe_sigcomm19, linefs_sosp21, fasst, drtmh, grappa_atc15, drtm}, we adopt a shared-record organization where transactions on any node may manipulate any records of the dataset. 
Nodes execute in a symmetric model~\cite{fasst, drtmh, drtm, farm_nsdi14}, where each node runs both client and server processes. 
Note that \name~is also compatible with other organizations, such as shared-nothing architectures~\cite{hstore_damon16, citus_sigmod21, memsql_vldb16, voltdb, calvin_sigmod12} and deterministic scheduling~\cite{caracal_sosp21, calvin_sigmod12}.
We adopt the widely used framework DBx1000~\cite{abyss_vldb14, taurus_vldb2020} as our codebase. The KVS uses either a hash table or a tree to store the key-value items. 
Taking hash index as an example, each key maps to a bucket, and the bucket is organized as a linked chain. 
A bucket item contains a pointer to the tuple header, which further points to the value. 
We bind only the records with \name~primitives, while other shared contents, including indexes and headers, remain with \vanilla~primitives.


\ifx\undefined\stale
\subsubsection{Transaction. }
We denote a set of tuples read and written by a transaction by the read set (\textit{RS}) and the write set (\textit{WS}). We assume that a transaction first reads the tuple it writes so that $RS \subseteq WS$. We use the four phase OCC as an example to illustrate how a transaction benefits from \name~primitives. 

\noindent \textbf{(1) Execution Phase. }The transaction begins execution by reading the header and the keys from the dataset. For a key in \textit{WS}, the worker thread allocates a buffer in the local memory and redirect writes to it. \textit{The isolation nature of \name-primitive's loosely coherent model allows the \textit{record field} accesses in this phase free of cross-node coherence issues hence exhibit lower latency. }

\noindent \textbf{(2) Validation Phase. } After accessing all tuples, the worker thread tries to lock each tuples in the \textit{WS}. \textit{This is done by standard single-node atomic operations, e.g. compare-and-swap (CAS) and fetch-and-add (FA), thanks to the strict cache coherence model on this field. }
If any key is locked or the version has changed from the first phase, the transaction will abort.

\noindent \textbf{(3) Logging Phase. } The worker thread poses the \textit{WS}'s key-value items and their versions to the buffer of helper threads and returns. We implement a batched logging scheme of taurus~\cite{taurus_vldb2020}. \textit{Logs are flushed to local file systems so that avoid intervening the shared memory. }

\noindent \textbf{(4) Commitment Phase. } 
In addition to copy WS's records to the shared locations, the worker thread should also call the GSync on the records' addresses and waits for an completion from the EP. 
\textit{The system could batch GSync calling from different transactions on the same node~\cite{fasst}, or hide the latency with co-routines~\cite{drtmh}. } After receiving EP's responses, the worker thread then increases the tuples' versions and unlocks the keys. 
\textit{The withdrawn writes would not cause any cross-node coherence issues. }


\fi




\ifx\stale\undefined
\begin{itemize}
    \item \textbf{Execute. } The transaction begins execution by reading the header and the keys from the dataset. For a key in \textit{WS}, the worker thread allocates a buffer in the local memory and redirect writes to it. The isolation nature of \name-primitive allows the data field reads in this phase without cross-node coherence issues. But in CXL-vanilla, a data field read may access remote caches if a transaction has modified it but not written it back.

    \item \textbf{Validate. } After accessing all tuples, the worker thread tries to lock each tuples in the \textit{WS}. This is done by standard single-node atomic operations, e.g. compare-and-swap (CAS) and fetch-and-add (FA), toward the headers bounded with CXL-vanilla primitives. If any key is locked or the version has changed from the first phase, the transaction will abort.

    \item \textbf{Log. } If validation succeeds, the worker thread can log the modified content to non-volatile storage. The worker thread poses the \textit{WS}'s key-value items and their versions to the buffer of helper threads and returns. We implement a batched logging scheme resembling taurus~\cite{taurus_vldb2020}. Each node adopts a helper thread to store logs in local file systems. The worker thread poses modified data to a buffer and returns, the helper thread checks the buffer and flush it at 1ms interval.  

    \item \textbf{Commit. } If logging succeeds, the worker thread updates the records in WS and explicitly synchronizes the content to other nodes. It first copies each tuples in WS from the buffer to the shared locations. This step advertises the changes to all local peer processors. To advertise other nodes, the transaction calls GSync on the the location and waits for an completion signal from the EP. After that, the worker thread increases the tuples' versions and unlocks the keys. 

    \item \textbf{Abort. } If the transaction aborts, it cleans up the temporal buffers, then cancels the updates on shared records with Wd (if has). Note that Wd is not used in OCC since writes are buffered, but the Wd is required in the pessimistic concurrency control policies such as two phase locking to withdraw writes. 
\end{itemize}
\fi

\subsection{Setups}  \label{subsec:setup}

\noindent \textbf{Benchmarks.} 
Following previous in-memory transaction processing works~\cite{dbx1000_dist_vldb17, drtm, drtmh, polyjuice_osdi21}, we adopt two OLTP tasks: 

\noindent (a) TPC-C~\cite{tpc-c}, the current industry standard for OLTP evaluation. 
Similar to previous works~\cite{dbx1000_dist_vldb17}, we adopt two (Payment and NewOrder) out of five transactions in our simulation as a default mixture (DM) since they account for 88\% of the TPC-C workload, and also tests NewOrder-only cases (NO). 

\noindent (b) YCSB~\cite{ycsb}, representative of large-scale cloud services. In this paper, we use a 16GB YCSB database containing a single table with 16 million records. Each tuple has a single primary key and 10 values, each with 20 bytes of randomly generated string data. 
A transaction accesses 16 records with reads or writes controlled by write ratio $w$, and follows the Zipfian distribution controlled by the skew factor $\theta$.





\begin{figure*}[t]
  \centering
  \includegraphics[width=0.93\textwidth]{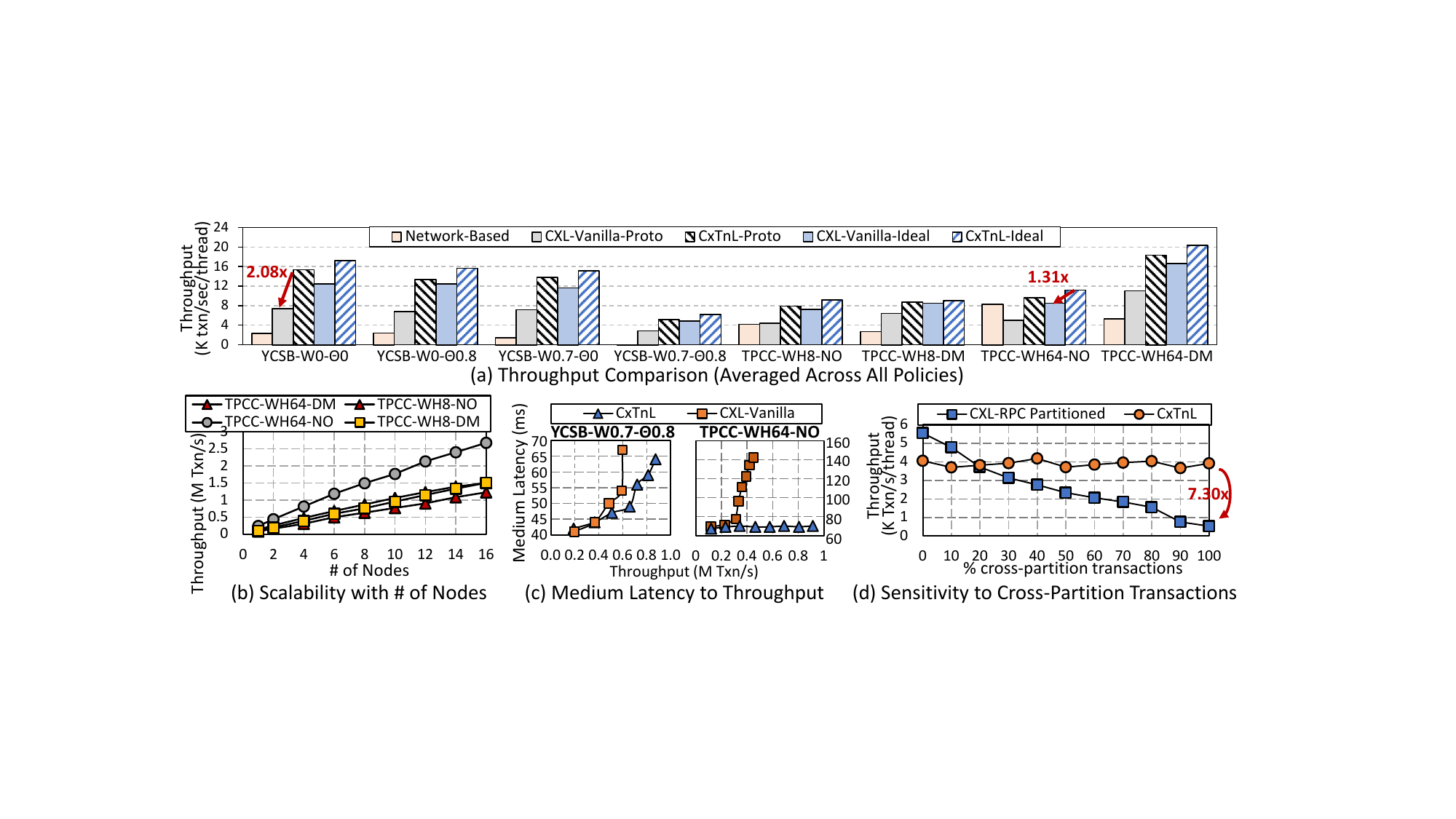}
  \caption{Overall Performance Comparison. }
  \label{fig:throughput}
\end{figure*}

\begin{table}[t]
\begin{minipage}{0.45\textwidth}
\caption{System Parameters and Configurations}\vspace{-5pt}
\label{tab:system-config}
\resizebox{\textwidth}{!}{%
\begin{tabular}{@{}|cl@{}}
\toprule
\multicolumn{2}{c}{\textbf{Host Processors}}                                                    \\ \midrule
\multicolumn{1}{c|}{Processor}     & 10-core @ 3.2GHz                                   \\ \midrule
\multicolumn{1}{c|}{L1I/L1D/L2/L3} & 32KB,4way / 32KB,8way / 2MB,8way / 1.875MB,16way                    \\ \midrule
\multicolumn{1}{c|}{Memory System} & 2x channel, 64GB, DDR4 @2666MHz                   \\ \midrule
\multicolumn{2}{c}{\textbf{CXL Configurations}}                                                      \\ \midrule
\multicolumn{1}{c|}{Connection}    & 32Gb/s/Lane, 8x Lanes (32GB/s) per node     \\ \midrule
\multicolumn{1}{c|}{Memory}        & \begin{tabular}[c]{@{}l@{}}DDR4 @2666MHz, 21.3GB/s/channel,\\ 8x channels (170.4GB/s in total)\end{tabular} \\ \midrule
\multicolumn{1}{c|}{Snoop Filter}  & 128K entries, 16 way               \\ \midrule
\multicolumn{2}{c}{\textbf{CxTnL Parameters}}                                                   \\ \midrule
\multicolumn{1}{c|}{VF/VBF}        & 512B per node,2 hash / 16KB per node,2 hash                            \\ \midrule
\multicolumn{1}{c|}{VAT} & \begin{tabular}[c]{@{}l@{}}1M entries/hash table, 2-way, 40 bit index, \\ resizing threshold=0.6/0.01, interval=10ms\end{tabular} \\ \bottomrule
\end{tabular}%
}
\end{minipage}
\end{table}

\noindent \textbf{Evaluation Methodology.}
We develop an FPGA-based CXL memory system, incorporating an Intel$^\circledR$ Sapphire-Rapids$^\texttt{TM}$ Xeon Gold 6430 CPU and an Intel$^\circledR$ Agilex$^\texttt{TM}$-7 I-Series FPGA configured as a CXL type-2 device (CXL 1.1). This configuration enables both \cxlmem~and \cxlcache~of the CXL IP~\cite{cxl_ip}. 
\rvs{
The CXL IP consists of a hardware component, implemented as an independent chiplet known as R-Tile~\cite{cxl_ip}, and a soft-encrypted wrapper located within the Programmable Logic (PL) domain, working at 400 MHz. The inter-chiplet communication between the hard and soft components of the CXL IP introduces a significant portion of the overall latency. 
The FPGA owns dual-channel DDR4-2666 memory, totaling 16GB in capacity. 
The memory system operates under Linux kernel v6.3. 


With this prototype, we evaluate two key performance features that determine the CXL protocol latency: the host-to-device \cxlmem~roundtrip latency and the device-to-host \cxlbi~roundtrip latency. To evaluate the first phase, we employ the approach used in prior studies~\cite{cxl_demystify, neomem} that adopts the standard memory latency checker~\cite{mlc} to generate loads and stores from the host CPU to the device memory. To mimic the snoop-filter behaviors as mentioned Figure~\ref{fig:motivation-over-coherent}, we implement a direct-map SF at the FPGA PL side, locating at \cxlmem~critical path and do udpates at each memory accesses. This number is known as C2M latency in Table~\ref{tab:latency}. 
For the second phase, we utilize \cxlcache~to emulate the datapath of \cxlbi~since it is not yet available in CXL 1.1 devices. \cxlbi~have similar protocol roundtrip constitution with \cxlcache, and it is introduced in 3.0 to avoid the circular channel dependence between \cxlcache~and \cxlmem~\cite{cxl-shortdoc, cxl-doc, cxl-paper}. 
We construct a test core on the FPGA's PL side to generate memory access requests. These requests are routed to the CXL IP in host-bias mode via the AXI bus, initiating \cxlcache~requests from the device to the host~\cite{cxl_ip, cxl-doc, cxl-shortdoc}.
The test core waits for the accessed data to be returned and records the total latency. 
To get the pure CXL link latency, we subtract the two aforementioned numbers with the time that PL test core spends on accessing the device memory without intervening CXL IP. We add the two subtraction results to get C2C latency.

Due to the deadlock issue from \cxlcache's channel dependency~\cite{cxl-shortdoc, cxl-doc, cxl-paper}, we are unable to conduct full system evaluation based on the prototype FPGA. We thus develop a pin-tool simulator based on Sniper~\cite{sniper} with around 2.5K LoC changes. We build a memory node that simulates the \name~and CXL-vanilla architecture based on the performance characteristics of the prototype. We change a few lines of DBx1000 to adapt to CTLib. 
We also assess the ideal ASIC performance in the simulator based on reports from chip vendors and IP providers~\cite{samsung, directcxl, tpp_asplos23, hash-ip}, as detailed in Table~\ref{tab:latency}. 
For network-based systems, we deploy the distribution version of DBx1000~\cite{dbx1000_dist_vldb17, abyss_vldb14} on 8 r320 nodes in CloudLab~\cite{cloudlab}. 
}



\noindent \textbf{Simulation Setups. }
Table~\ref{tab:system-config} summarizes the default system configurations. Each node equips with an OoO CPU with the Intel Xeon Gold 6430 architecture (Sapphire Rapids series). For CXL architecture, we equip each node with an P2P CXL $\times$8 connection with EP. The EP manages shared memory in 8 channels with interleaved addresses in cacheline granularity. The EP manages an 16 way on-chip SF with 128K-entries. For \name~architecture, each CTHW implements VF and VBF via 512B and 16KB bloom filters with 2 hash functions by default. 
The view shim indicator takes 56 bits with 4 bit node id and 48 bit EMS address. The most significant 16 bits index the on-chip LUT, the following 40 bits calculates the cuckoo hashes. 



\begin{figure}[t]
  \centering
  \includegraphics[width=0.47\textwidth]{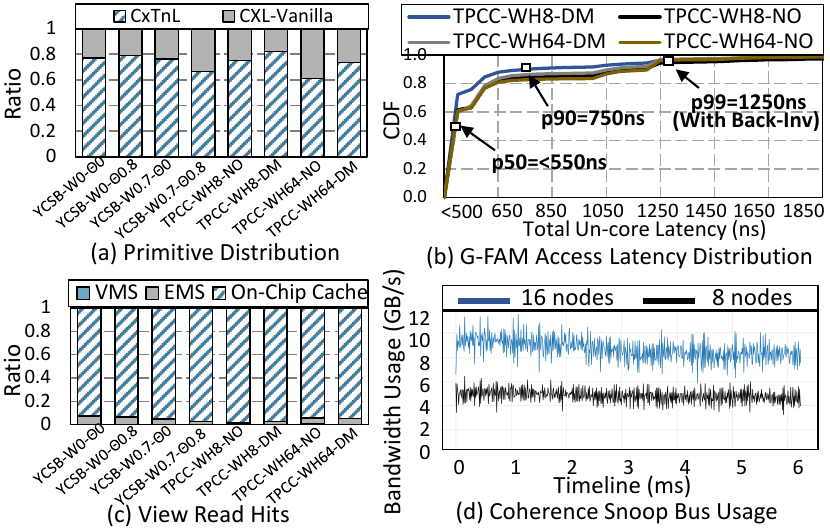}
  \caption{Detailed Analysis of \name.}
  \label{fig:dram_hist}
\end{figure}

\subsection{Main Results}

Figure~\ref{fig:throughput}~(a) presents a performance comparison between our proposed \name~and baseline primitives, with throughput averaged across all transaction control policies in the benchmark to neutralize algorithmic effects. The results highlight \name's consistent performance gains over CXL-vanilla systems, achieving a 1.36x improvement in ideal ASIC setups and 2.08x in prototype setups. Specifically, \name~exhibits greater enhancements in the memory-insensitive YCSB benchmarks (93\% on average) compared to the TPCC benchmarks (77\% on average), due to TPCC's higher computation burden. Moreover, \name~outstrips traditional networking-based architectures by 6.47x, leveraging the superior performance of the CXL protocol while circumventing the inefficiencies of its coherence model.

Figure~\ref{fig:throughput}~(b) shows the throughput with an increasing number of nodes, each dedicating 8 worker threads per node. It illustrates \name's good scalability with the number of sharing nodes.  
Figure~\ref{fig:throughput}~(c) shows the throughput-to-median latency relationship of \name~and \vanilla. We vary the load by increasing the number of worker threads per node from 1 to 8 while keeping 8 nodes. \name~is able to achieve 0.87M txn/sec with 59ms median latency in the highly contended YCSB workload, which is notably better than CXL vanilla with 0.59M txn/sec and 67ms median latency. On the less contended TPCC benchmarks, \name~achieves 0.92M txn/sec with stable median latency. 

Figure~\ref{fig:throughput}~(d) shows the throughput of running all new-order transactions on \name~and CXL-RPCs~\cite{hydrarpc_atc24, partial_sosp23} as we increase the number of cross-partition transactions. To make an apple-to-apple comparison, we adopt the distributed version of DBx1000~\cite{dbx1000_dist_vldb17} and store shared tuples at G-FAM, with each node exclusively managing 8 warehouses of TPC-C. We adopt SOTA CXL-shared memory RPC, i.e., HydraRPCs~\cite{hydrarpc_atc24}, which achieves 1.47us avg. latency by passing references instead of data. 
We test on 4 nodes, each with 8 worker threads. The CXL-RPC-based shared-nothing architecture is the optimal solution for perfectly partitionable workloads due to minimized cross-node coherence~\cite{hekaton_sigmod13}. However, when roughly 20\% of transactions touch multiple partitions, the throughput of the CXL-RPC architecture drops below \name.



\subsection{Detailed analysis of \name}

\begin{figure*}[t]
  \centering
  \includegraphics[width=0.90\textwidth]{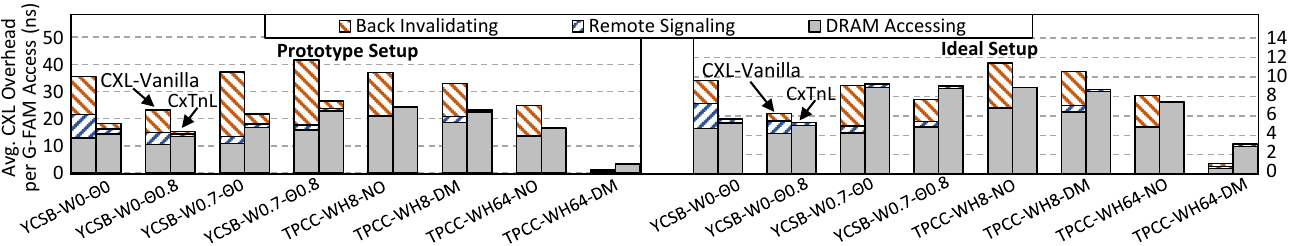}
  \caption{CXL-related overhead breakdown. }
  \label{fig:overhead_breakdown}
\end{figure*}

\noindent \textbf{CXL Overhead Breakdown.} Figure~\ref{fig:overhead_breakdown} reports the latency breakdown on G-FAM accesses with the configuration of the SILO algorithm, no logging, and a hash index. We break down the time into DRAM accessing time, which includes the time spent on the CXL IP and DRAM controller, remote signaling time, and snoop filter querying and back invalidation time. 
The numbers are averaged across all loads and stores towards the G-FAM address segment, including accesses that hit local caches. 
\name~reduces the average remote signaling cost by 65.9\% in the prototype and 70.8\% at the ASIC. 
\name~also reduces back invalidation overheads by 87.1\% and 96.5\% at maximum, respectively, due to the reduced SF conflicts and queries on records. 
Despite the endpoint logic increasing by 34.6\% and 44.4\% due to \name-introduced overheads, the overall cost reduces by 40.2\% and 23.7\%, respectively. 

\noindent \textbf{Primitive Distribution.}
Figure~\ref{fig:dram_hist}~(a) shows that 73.8\% of G-FAM accesses are \name~primitives, while only 26.2\% are \vanilla~for synchronization. Note that the fraction of \name~primitives could be larger in real-world applications due to larger record sizes and fine-grained primitive binds. 

\noindent \textbf{L-Ld/L-St Overheads.}
We further test the overheads of the \name~architecture. 
Figure~\ref{fig:dram_hist}~(b) illustrates the distribution of un-core latency for accesses that actually touch the G-FAM node. It shows that more than 50\% of G-FAM accesses take less than 500ns, and 90\% of accesses are bounded within 750ns. Given the zero-load \cxlmem~latency is around 480ns, the overhead of the view translation flow is limited to 270ns, which is 56.2\% relative to zero-load \cxlmem~latency. The p99 latency reaches 1250ns since it takes the CXL-vanilla primitive and causes back-invalidation. 

\begin{figure}[t]
  \centering
  \includegraphics[width=0.475\textwidth]{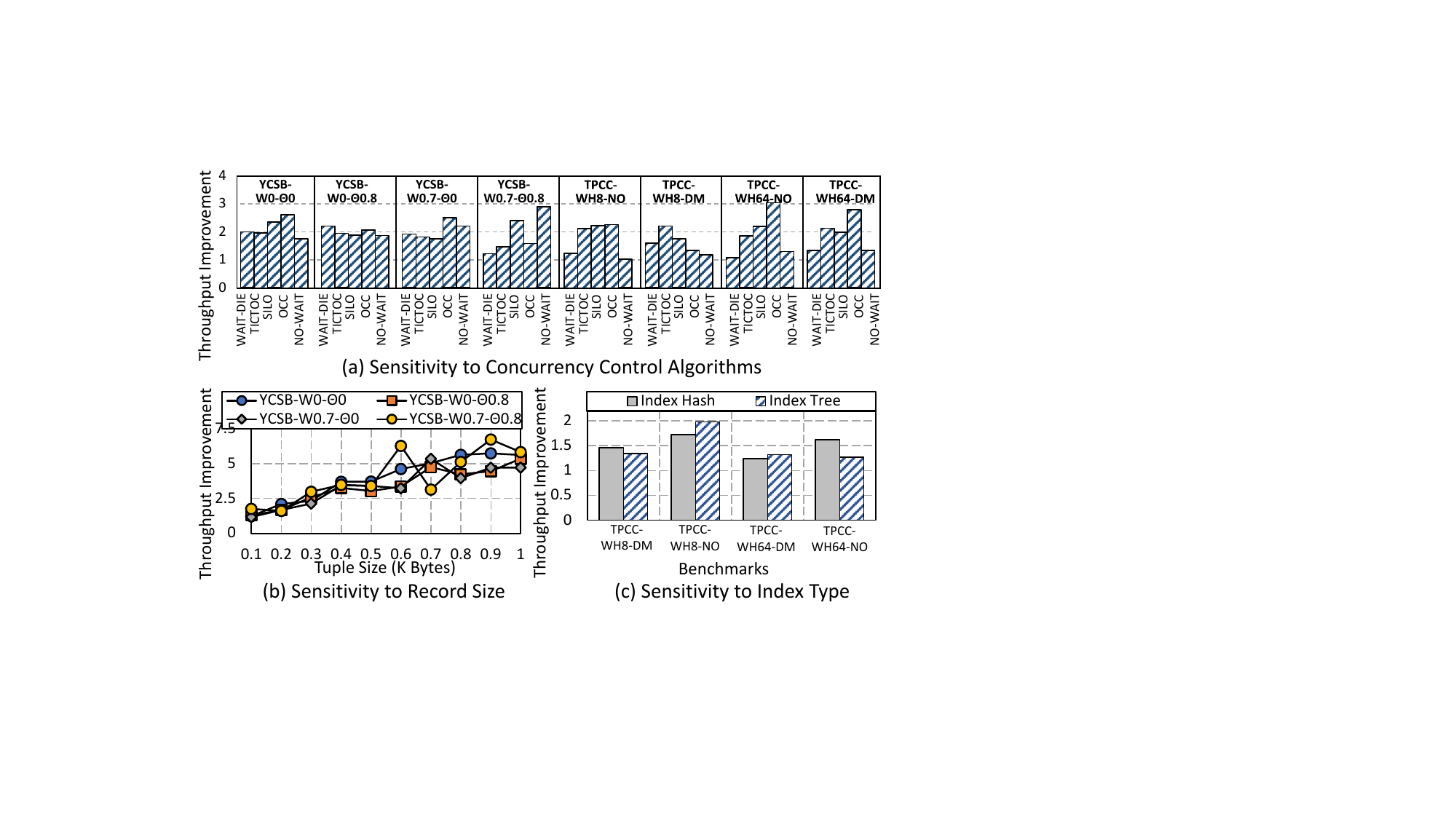}
  \caption{Generality Tests}
  \label{fig:record_sensitivity}
\end{figure}

\noindent \textbf{View Type Distribution.}
Figure~\ref{fig:dram_hist}~(c) shows the usage of VMS. It indicates that almost all view accesses are served by processors' caches and less than 0.2\% of views overflow to the VMS, thanks to the recent trend in processor design, which enlarges available on-chip caches by increasing L2 caches. Moreover, the VMS filter efficiently sifts out 96.1\% of VAT queries that might involve additional overheads. 

\noindent \textbf{Coherence Traffic.}
Figure~\ref{fig:dram_hist}~(d) shows the usage of internal bandwidth between CTHWs. In a 16-node scenario, the cross-node view synchronization traffic only takes up 8.54 GB/s at p99 usage. The bus usage is always under 12\% for a 512-bit bus working at 1GHz. This shows that the essential cross-node coherence traffic is in fact low, and our loosely coherent \name~primitive successfully exploits this.



\subsection{Generality to Software Policies}    \label{subsec:generality}

\noindent \textbf{Concurrency Control Sensitivity. } 
To demonstrate the wide compatibility to the wide range use cases, 
We first investigate the impact of transaction processing policies on system performance at Figure~\ref{fig:record_sensitivity}~(a). 
\name~exhibits broad performance improvement with average 1.90x on all algorithms. The OCC~\cite{occ_tbs81} and SILO~\cite{silo_sosp13} gains the highest speedup by the factor of 2.26x and 2.06x due to their speculative execution that always access the entire read/write set despite of aborts. 2PL-variant policies such as NO\_WAIT and WAIT\_DIE~\cite{2pl} gain lower speedups, i.e. 30.7\% and 20.3\% on average at TPC-C benchmarks, since they are bottlenecked by the lock managers at most cases~\cite{abyss_vldb14}.


\noindent \textbf{Record Size Sensitivity. } 
As Figure~\ref{fig:record_sensitivity}~(b) shows, \name~achieves higher speedup on the large record size, since it directly reduce the coherence maintenance cost on record memory accesses. This figure adopts a SILO concurrency control at the FPGA setup, with the fixed 8 byte metadata~\cite{silo_sosp13}. By increasing the YCSB record size from 100B to 1KB per tuple, the speedup increases from 1.22x to 5.81x accordingly. This result demonstrates \name's opportunity for the applications with large record size.

\noindent \textbf{Index Sensitivity. } 
Figure~\ref{fig:record_sensitivity}~(c) shows \name's generality to hash indexes and binary tree indexes~\cite{abyss_vldb14, masstree} in TPC-C benchmarks. Despite the current index design is bounded to original CXL vanilla primitives that could not directly benefit from \name~primitives, we still achieve 1.72x and 1.98x performance improvement on hash and binary trees respectively, since \name~avoid the SF eviction traffic from tracking large records. 



\begin{figure}[t]
  \centering
  \includegraphics[width=0.485\textwidth]{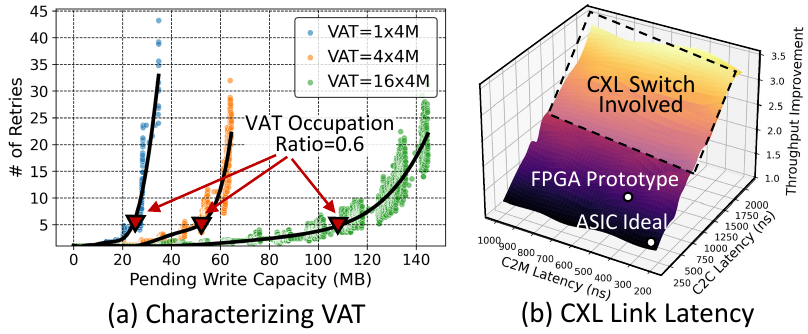}
  \caption{\rvs{Scalability Tests}}
  \label{fig:latency_sweep}
\end{figure}

\rvs{
\subsection{Scalability Tests}  \label{subsec:latency_sweep}
\noindent \textbf{VAT Scalability.}
Figure.~\ref{fig:latency_sweep}~(a) illustrates the relationship between the total pending write capacity of transactions and the number of VAT retries required for successful element insertion. We build a single-node synthesized-benchmark that generates random 64-byte L-St requests until the VAT reaches the maximum retry threshold (set as 64). The host CPU owns 32 cores and each has 3.85MB on-chip caches and adopts random eviction strategy on cache conflict. 
A cuckoo hash table occupies 4MB DRAM accommodating 512K entries. 
The results demonstrate that when the number of pending writes is relatively small (less than 20 MB), the insertion is often successful on the first attempt or requires only a few retries. As the write capacity increases, the insertion performance deteriorates. But we could keep the number of retries under 6 by  maintaining the VAT occupation under 0.6. Under our default setup, the system could support the application with the write set capacity of 27MB, 52MB, and 112MB per node with the configurations using 1, 4, and 16 cuckoo hash tables, respectively. 

Default VF and VBF can hold about 1.4K and 53K items at about 25\% false-positive rate. Note that the default configuration of the CTHW is based on the general OLTP workloads, to optimize for applications with larger pending write sets, users could enlarge the default size or adopt pessimistic concurrency controls such as 2PL variants~\cite{abyss_vldb14}. 


\noindent \textbf{Link Latency.}
The CXL specification 2.0 introduces the CXL switches to facilitate connectivity in large-scale clusters. However, incorporating CXL switches, along with necessary retimers, significantly increases the link latency. 
To study the \name~sensitivity on link latency, we conduct a what-if analysis by adjusting C2C and C2M latency from hundreds of nanoseconds to several microseconds. 
Figure.~\ref{fig:latency_sweep}~(b) depicts the throughput improvement of \name~over the CXL-vanilla baseline. We test SILO algorithm on the TPC-C benchmark with 8 nodes and 8 threads each. We could tell that CXL link latency positively related to performance improvement due to its direct impact on the C2C overheads. \name~can achieve a throughput improvement of up to 3.41x when the CXL link takes near 1ms. 

These findings indicate that the advantages of \name~grow as the cluster size increases. However, we suggest that the sweetpoint scale for \name~is when the C2C and C2M roundtrip latency is below one microsecond, which is typically a rack containing 8-16 nodes with direct connection to shared memory~\cite{pond}. Because within the host processors, CXL accessing datapath share many latency hiding modules, such as store buffers and miss-status handling registers, with main memory accessing datapath. Excessive latency in CXL memory access may overwhelm these modules and effect main memory accessing performance.
}




\section{Discussion}

\rvs{
\subsection{Scaling issues beyond a rack}
As we discussed in last section, we believe that a rack with 8-16 nodes is the sweetpoint scale for \name, and 
beyond a rack, we anticipate the need for a heterogeneous interconnection approach, where intra-rack communication leverages CxTnL to enable efficient memory sharing, and inter-rack communication relies on traditional networks such as Ethernet or RDMA. The reasons are twofold. 

\noindent \textbf{Broadcasting limitations.} \name~adopts a broadcasting-based coherence design to avoid SF bottlenecks, but this scheme does not scale well with an increasing number of nodes due to the flooding snooping traffic. Within a rack, the broadcasting field with up to 16 nodes could be efficiently supported by an in-device hardware bus, while scaling beyond a rack requires multiple CxTnL devices working in a distributed way due to I/O port limitations~\cite{pond}. This introduces costly cross-device communication for broadcasting. 

\noindent \textbf{Node Failure Issues.} Fault tolerance is crucial in large-scale transaction processing systems. To maximize the efficiency of the critical path, \name~hardware does not implement failure recovery mechanisms. CxTnL will stops the world and raises a hardware error when probes a node failure, and relies on the transaction processing frameworks to ensure data consistency. Consequently, the scale of \name's becomes the blast radius of one node failure. By limiting \name~within a rack, a system can isolate blast regions using network-based fault-tolerance protocols such as 2PC.



\subsection{Related works}


\noindent \textbf{Weak Consistency Models.} A consistency model defines the contract between hardware architecture and software developers, specifying the expected behavior of a memory load. For software developers, the most intuitive model is sequential consistency~\cite{lamport_tc79}, which maintains program order and write atomicity for parallel processes. To boost hardware performance, prior works~\cite{rvweak_pact17, rvweak_isca18, spandex_asplos18, lrcgpu_micro16, lrcdsm_isca92} have explored a variety of weaker consistency models by relaxing certain guarantees of sequential consistency. \name~adopts the consistency model similar to release consistency~\cite{threadmarks_tc94, munin_ppopp90, txcache_osdi10, lrcgpu_micro16} which preserves a partial order of memory accesses. \name~details how this consistency model benefits CXL-based memory sharing and provides holistic hardware and software design.

\noindent \textbf{Distributed Transaction Processing.}
Distributed transaction processing frameworks can be categorized into two main architectures: shared-nothing and shared-data. For decades, the shared-nothing architecture has dominated distributed database management systems, primarily due to its efficiency in minimizing network communication~\cite{hstore_damon16, citus_sigmod21, memsql_vldb16, voltdb, calvin_sigmod12, dbx1000_dist_vldb17}. 
The shared-data architecture allows transactions to access and modify data across any node within the system~\cite{mtcp_nsdi14, xenic_sosp21, ipipe_sigcomm19, linefs_sosp21, fasst, drtmh, grappa_atc15, mtcp_nsdi14, drtm}. \name~design adheres to shared-data architectures but also offers potential benefits to shared-nothing architectures by optimizing conventional network communication primitives, such as HydraRPC~\cite{hydrarpc_atc24}. 
}

\section{Conclusion}
The emergence of advanced interconnection techniques CXL enables us to rethink the distributed transactions. To overcome the performance drawbacks of CXL's vanilla coherence model, this work proposes \name, a software-hardware co-designed system that implements a novel hybrid coherence primitive tailored to the loosely coherent nature of transaction processing systems. Evaluations on OLTP benchmarks demonstrate the benefits of \name, which achieves 2.08x higher throughput than \vanilla~primitives and exhibits unified performance improvements on various transaction processing policies. 

\end{sloppypar}

\bibliographystyle{IEEEtranS}
\bibliography{references}

\end{document}